\documentclass[journal=jacsat,manuscript=article]{achemso}

\usepackage[version=3]{mhchem} % Formula subscripts using \ce{}
\usepackage{amsmath}
\usepackage{amssymb}
\usepackage{mathrsfs}
\usepackage{amsfonts}
\usepackage{bm}
\usepackage{dsfont}
\usepackage{verbatim}
\usepackage{braket}
\usepackage{hhline}   % for fancy table lines
\usepackage[absolute]{textpos}
\usepackage{xcolor}
\usepackage{mathrsfs}
\usepackage{nicefrac,xfrac}
\usepackage{subcaption}
\usepackage{siunitx}
\SectionNumbersOn
\usepackage{achemso}
\setkeys{acs}{maxauthors = 0}
\author{Yuchen Wang}
\affiliation{Department of Chemistry, Department of Physics and Purdue Quantum Science and Engineering Institute, Purdue University, West Lafayette, Indiana 47907, USA}
\author{Ellen Mulvihill}
\affiliation{Department of Chemistry, Yale Quantum Institute, Yale University, New Haven, CT 06511, USA }
\author{Zixuan Hu}
\affiliation{Department of Chemistry, Department of Physics and Purdue Quantum Science and Engineering Institute, Purdue University, West Lafayette, Indiana 47907, USA}
\author{Ningyi Lyu}
\affiliation{Department of Chemistry, Yale Quantum Institute, Yale University, New Haven, CT 06511, USA }
\author{Saurabh Shivpuje}
\affiliation{Department of Chemistry, Department of Physics and Purdue Quantum Science and Engineering Institute, Purdue University, West Lafayette, Indiana 47907, USA}
\author{Yudan Liu}
\affiliation{Department of Chemistry, University of Michigan, Ann Arbor, MI 48109, USA}
\author{Micheline B. Soley}
\affiliation{Department of Chemistry, Yale Quantum Institute, Yale University, New Haven, CT 06511, USA }
\alsoaffiliation{Department of Chemistry,  Department of Physics, University of Wisconsin-Madison, Madison, WI 53706, USA}
\author{Eitan Geva}
\affiliation{Department of Chemistry, University of Michigan, Ann Arbor, MI 48109, USA}
\email{eitan@umich.edu}
\author{Victor S. Batista}
\affiliation{Department of Chemistry, Yale Quantum Institute, Yale University, New Haven, CT 06511, USA }
\email{victor.batista@yale.edu}
\author{Sabre Kais}
\affiliation{Department of Chemistry, Department of Physics and Purdue Quantum Science and Engineering Institute, Purdue University, West Lafayette, Indiana 47907, USA}
\email{kais@purdue.edu}
\title[An \textsf{achemso} demo]
  {Simulating Open Quantum System Dynamics on NISQ Computers with Generalized Quantum Master Equations}

\begin{document}

\begin{tocentry}
\includegraphics[width=\textwidth]{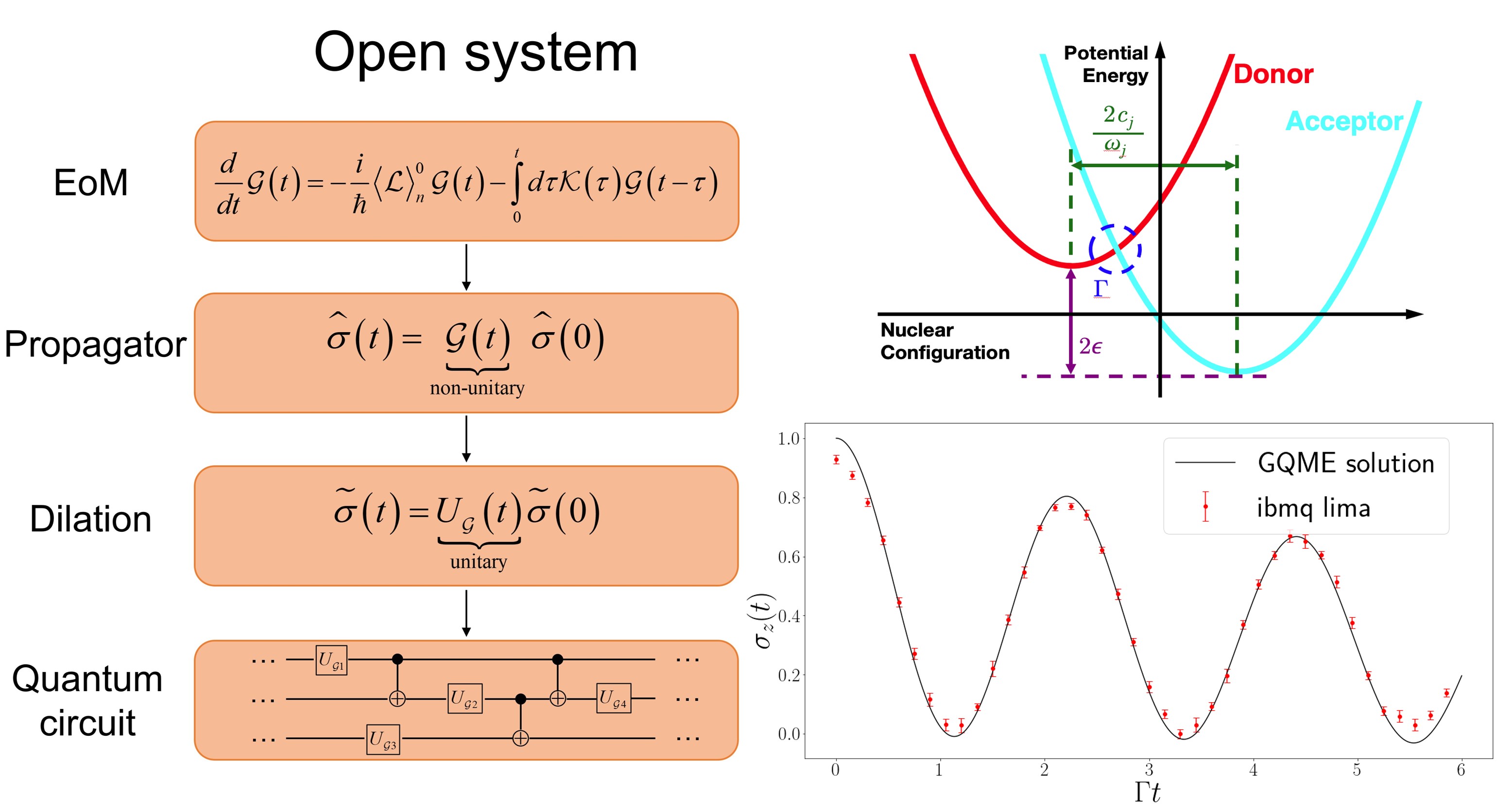}
\end{tocentry}

\begin{abstract}
  We present a quantum algorithm based on the Generalized Quantum Master Equation (GQME) approach to simulate open quantum system dynamics on noisy intermediate-scale quantum (NISQ) computers. This approach overcomes the limitations of the Lindblad equation, which assumes weak system-bath coupling and Markovity, by providing a rigorous derivation of the equations of motion for any subset of elements of the reduced density matrix. The memory kernel resulting from the effect of the remaining degrees of freedom is used as input to calculate the corresponding non-unitary propagator. We demonstrate how the Sz.-Nagy dilation theorem can be employed to transform the non-unitary propagator into a unitary one in a higher-dimensional Hilbert space, which can then be implemented on quantum circuits of NISQ computers. We validate our quantum algorithm as applied to the spin-boson benchmark model by analyzing the impact of the quantum circuit depth on the accuracy of the results when the subset is limited to the diagonal elements of the reduced density matrix.  Our findings demonstrate that our approach yields reliable results on NISQ IBM computers.
\end{abstract}

\section{Introduction}
\label{sec:introduction}
\begin{figure}[h]
\centering
\includegraphics[width=0.45\textwidth]{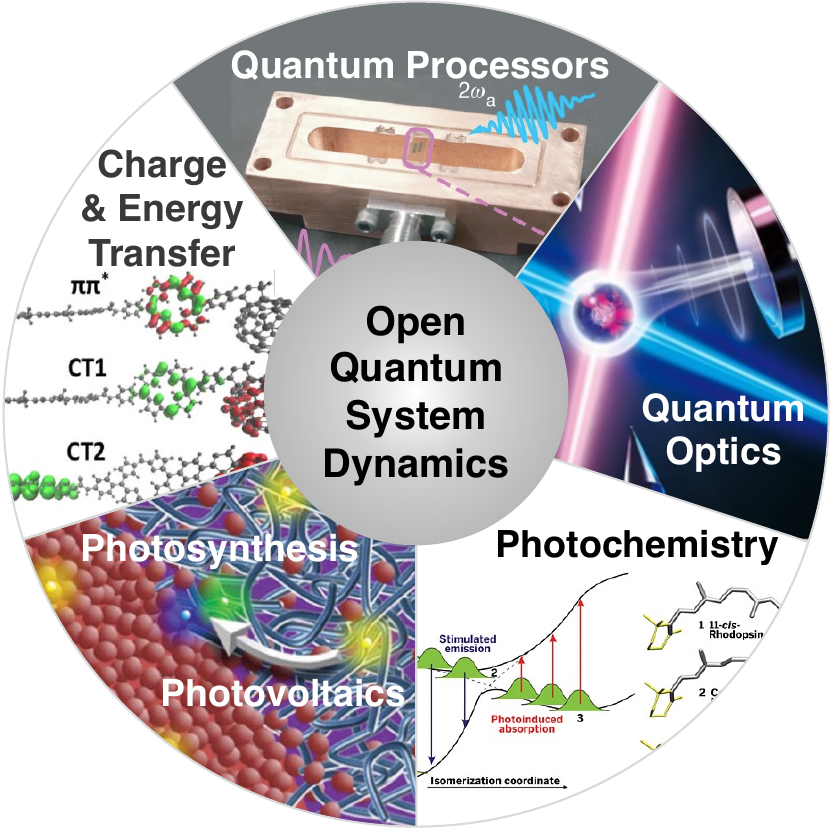}
\caption{The simulation of open quantum system dynamics is central to many science and engineering disciplines (a few examples are showcased in the figure).
%Open quantum system dynamics therefore serves as a unifying interdisciplinary platform that bridges the gap between theory and experiment.
%Important disciplines that involve open quantum system dynamics.
}
\label{fig:openSystems}
\end{figure}

Simulations of open quantum systems have become essential for studying the dynamics of quantum systems in the condensed phase, allowing for the inclusion of dissipative effects from the environment which are critical for accurate simulations. These powerful computational tools have enabled a wide range of studies, from chemical and physical processes to excited state lifetimes, spectral diffusion and line-broadening, across multiple fields of research, including physical chemistry, molecular physics, condensed-phase physics, nanoscience, molecular electronics, quantum optics, nonequilibrium statistical mechanics, spectroscopy and quantum information science.~\cite{wangsness53,redfield57,bloch57,haake73,yoon75,lindblad76,gorini76b,oppenheim77,alicki87,grabert82,kubo83,laird91a,vankampen92,pollard96,kohen97,kosloff97,cao97,shi03f,baiz11,zwanzig01,may00,nitzan06,breuer02,jang21,lai21,chen22,mulvihill19a,mulvihill21a,mulvihill21b,mulvihill22} Examples of open quantum system dynamics include energy and charge transfer, dephasing, vibrational relaxation, nonadiabatic dynamics and photochemistry (see Fig.~\ref{fig:openSystems}). By harnessing the power of open quantum system simulations, we can bridge the gap between theory and experiment, providing insight into various complex phenomena in a variety of light-induced physical and chemical processes, including photoinduced processes such as energy and charge transfer, vibronic relaxation, dephasing, and nonadiabatic dynamics.~\cite{xu94, ishizaki12, liddell97, liddell02,bredas04,rizzi08, tian11, mishra09, feldt10, zhao12, lee13, lee14a, hu2018dark, hu2018double, breuer02, nitzan06, daley14, wiseman09, mulvihill21a, brian21, dan22,kais2014quantum}. 

Recent advances in quantum computing have enabled the development of numerous algorithms for electronic structure calculations,~\cite{peruzzo2014variational,o2016scalable,xia2017electronic,xia2018quantum} and simulations of quantum dynamics of closed quantum systems~\cite{wiebe2011simulating,ollitrault2021molecular,yao2021adaptive,tagliacozzo2022optimal}. 
However, relatively few studies have explored the simulation of open quantum system dynamics~\cite{lloyd2001engineering,wang2011quantum,wang2013solovay,wei2016duality,kliesch2011dissipative,sweke2015universal,schlimgen2021quantum,zhang2022quantum,guimaraes2023noise,rossini2023single}. 
These studies have been mostly based on Lindblad-type quantum master equations (QMEs) which ensure complete positivity and conservation of probability but rely on the Markov and Born approximations in the system-bath weak coupling limit~\cite{alicki87}. With the aim of developing a more general approach, here we introduce a quantum algorithm based on the Generalized Quantum Master Equation (GQME), which corresponds to the formally exact equation of motion (EoM) for an open quantum system. 

A major challenge facing the quantum simulation of open quantum system dynamics is the fact that the time evolution operators are non-unitary whereas quantum gates are unitary.
To this end, we have previously developed a quantum algorithm for open quantum dynamics based on the Sz.-Nagy unitary dilation theorem, which converts non-unitary operators into unitary operators in an extended Hilbert space. This algorithm was originally applied to simulating a Markovian two-level model  on IBM quantum computers.\cite{hu2020quantum} Later, the same method was applied to simulating the non-Markovian Jaynes-Cummings model on IBM quantum computers.\cite{head2021capturing} In a recent work, the same Lindblad-QME-based quantum algorithm was applied to simulate the dynamics of the Fenna-Matthews-Olson complex, which includes five quantum states and seven elementary physical processes.\cite{hu2022general} Thus far, this quantum algorithm has been used to simulate the dynamics of open quantum systems described by the operator sum representation or Lindblad-type QMEs. 

\begin{figure}[h]
\centering
\includegraphics[width=\textwidth]
{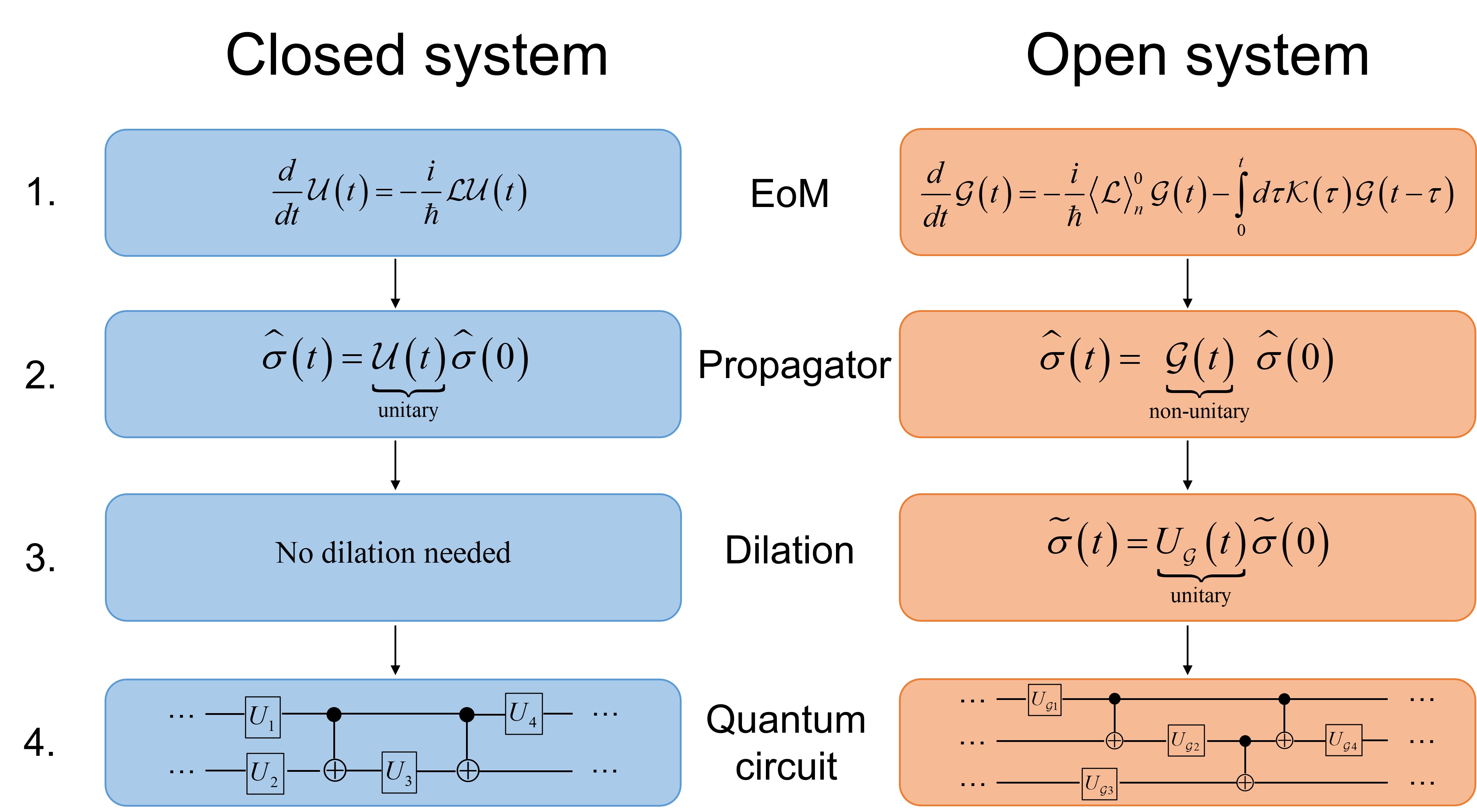}
\caption{A comparison of the workflows for simulating the dynamics of  a  closed quantum system governed by the quantum Liouville equation vs.~an open quantum system governed by the GQME. 
1. The EoM is established; 2. the time evolution superoperator is generated from the EoM; 3. A unitary dilation is required in order to convert the GQME-based non-unitary time evolution superoperator into a unitary superoperator in an extended Hilbert space; 4. Translation of the unitary matrix 
into a quantum gate sequence.}
	\label{fig:workflow}
\end{figure}

However, these approaches are not entirely general: the Lindblad QME used in Ref.~\citenum{hu2022general} relies on several restrictive approximations, including Markovian dynamics, and the ensemble of Lindbladian trajectories method in Ref.~\citenum{head2021capturing}, while capable of describing non-Markovian dynamics, involves user selection of {\em ad-hoc} system-bath parameters, therefore limiting the range of applications.
Furthermore, while the operator sum representation of open quantum system dynamics is general, it requires knowledge of the Kraus operators, which to the best of our knowledge are only known in closed form for systems whose dynamics can be described by Lindblad-type QMEs.

Extending the range of quantum simulation of open quantum systems  therefore calls for formulating the dynamics within a less restrictive theoretical framework.  The GQME formalism introduced by Nakajima \cite{nakajima58} and Zwanzig \cite{zwanzig60b} represents such a general framework since the GQME 
corresponds to the formally exact EoM of the open quantum system, as opposed to the Lindblad-type QMEs which correspond to approximate EoMs of the open quantum system.

A comparison of the workflows for simulating the dynamics of  a  closed quantum system governed by the quantum Liouville equation vs.  an open quantum system governed by the GQME is shown in Fig.~\ref{fig:workflow}. The derivation of the GQME involves projecting out the bath degrees of freedom (DOF) to obtain the EoM of the system's reduced density matrix,
or a subset of its elements. 
Within this EoM, which is referred to as the GQME, the memory kernel superoperator, $\cal{K} (\tau)$,   accounts for the main impact of the bath on the system's dynamics. Thus, the GQME replaces the Liouville equation as the formally exact EoM of the system when we transition from a closed quantum system to an open quantum system, with the memory kernel playing a similar role in the open system to that of the Hamiltonian or Liouvillian in the closed system.

In this work, we develop a GQME-based quantum algorithm for simulating the dynamics of an open quantum system. To this end, we develop a protocol for obtaining the non-unitary time evolution superoperator, or popagator, from the memory kernel. Then the Sz.-Nagy unitary dilation theorem is used to convert the GQME-based non-unitary 
%time evolution superoperator 
propagator into a unitary superoperator in an extended Hilbert space. Given this dilated and now unitary time evolution superoperator and the initial state of the system, we can evolve the dynamics for any open quantum system on quantum computers. 

Given the fact that the GQME is the exact EoM of the open quantum system, this quantum algorithm greatly extends the range of possible systems that can be simulated on a quantum computer, including complex non-Markovian photosynthetic and photovoltaic systems\cite{pfalzgraff19, mulvihill21a}, molecular electronics\cite{dan22}, linear and nonlinear spectroscopy\cite{fetherolf17}, systems with inter-system crossing\cite{fay19}, and conical intersections\cite{schile19}. Thus, this GQME-based quantum algorithm provides an essentially universal protocol for simulating open quantum system dynamics on quantum computing platforms. Given a powerful enough quantum computer, this algorithm opens the door for simulating open quantum system dynamics of large and complex molecular systems, which are currently beyond the reach of classical computers.

\section{Methods}

\subsection{GQME-based propagators 
%for the reduced density matrix of the quantum open system
%time evolution superoperator
}
\label{subsec:GQME}

In this section, we outline our approach for calculating the GQME-based non-unitary propagator for the reduced density matrix of the open quantum system (see Eq. (\ref{eq:G_nonunitary})). 
The analogous procedure for calculating the non-unitary propagator for a subset of the 
reduced density matrix elements 
%based on reduced-dimensionality GQMEs 
is outlined in Sec. \ref{subsec:red_dim_gqme}.

Previously developed quantum algorithms for open system dynamics involved mapping Lindblad operators to Kraus operators before using the Sz.-Nagy dilation theorem to reach a unitary quantum algorithm\cite{hu2020quantum, head2021capturing, hu2022general}. While useful for many systems, these methods are either Markovian\cite{hu2020quantum, hu2022general} or involve user selection of {\em ad-hoc} system-bath parameters\cite{head2021capturing}, therefore limiting the range of applications. In this paper, we introduce a method based on the GQME, %an {\em ab initio}, 
a formally exact EoM for the dynamics of an open quantum system.
%as opposed to the Lindblad equation. 
Instead of casting the non-unitary propagator in terms of Kraus operators and dilating them, this method uses the GQME to obtain the system's time evolution superoperator, or propagator, ${\cal G}(t)$, and perform the dilation on it to obtain a unitary quantum algorithm. This subsection describes the first step in the workflow outlined in Fig.~\ref{fig:workflow}, namely obtaining the time evolution superoperator of an open quantum system starting from its formally exact EoM in GQME form.

For the sake of concreteness, we will focus on molecular systems with an overall Hamiltonian of the following commonly encountered form:
\begin{equation}
\hat{H} = \sum_{j=1}^{N_e} \hat{H}_j | j \rangle \langle j| + \sum_{\substack{j,k=1 \\ k \neq j}}^{N_e} \hat{V}_{jk} | j \rangle \langle k|~~ 
\label{eq:genH}
\end{equation} 
and an overall system initial state of the following commonly assumed single-product form:
\begin{equation}
\hat{\rho} (0) = \hat{\rho}_n (0) \otimes \hat{\sigma} (0)~~.
\label{init_stat}  
\end{equation}
With this assumption,
the evolution is guaranteed to be described by a Completely Positive (CP) map~\cite{rivas2012open,wang2013solovay}. It should be noted that the GQME approach is not limited to this form of Hamiltonian and initial state and that the choice to focus on them is solely motivated by clarity of presentation 
 and the wide range of applications based on an Hamiltonian and an initial state of this form.
The system and bath in this case correspond to the electronic and nuclear DOF, respectively. 
In Eqs.~(\ref{eq:genH}) and (\ref{init_stat}),
$\hat{H}_j = \hat{\bf P}^2/2 + V_j \left( \hat{\bf R} \right)$ is the nuclear Hamiltonian when the system is in the diabatic electronic state $| j \rangle$, with the index $j$ running over the $N_e$ electronic states; $\hat{\bf R} = \left( \hat{R}_1,..., \hat{R}_{N_n} \right)$ and $\hat{\bf P} = \left( \hat{P}_1,..., \hat{P}_{N_n} \right)$ are the mass-weighted position and momentum operators of the ${N_n} \gg 1$ nuclear DOF, respectively; $\left\{ \hat{V}_{jk} | j \neq k \right\}$ are the coupling terms between electronic states (which can be either nuclear operators or constants); 
and $\hat{\rho}_n (0)$
and $\hat{\sigma}(0)$
are the reduced density operators that describe the initial states of the nuclear (bath) and electronic (system) DOF, respectively.
Throughout this paper, boldfaced variables, e.g., ${\bf A}$, indicate vector quantities; a hat over a variable, e.g., $\hat{B}$, indicates an operator quantity; and calligraphic font, e.g., ${\cal L}$, indicates a superoperator.

Using projection operator techniques, one can then derive the following formally exact EoM, or GQME, for the reduced electronic density operator, $\hat{\sigma} (t)$
\cite{mulvihill19a,mulvihill21a,mulvihill21b,mulvihill22}:
\begin{equation}
\frac{d}{dt} \hat{\sigma} (t) = -\frac{i}{\hbar} \langle \mathcal{L} \rangle_n^0 \hat{\sigma} (t) - \int_0^t d\tau\ \mathcal{K}(\tau)\hat{\sigma} (t - \tau)~~.
\label{eq:mGQME}
\end{equation}
The open quantum system dynamics of the reduced electronic density matrix described by this GQME is generated by the two terms on the R.H.S. of Eq.~\eqref{eq:mGQME}. The first term is given in terms of the projected overall system Liouvillian $\langle {\cal L} \rangle_n^0 \equiv \text{Tr}_n \left\{ \hat{\rho}_n (0) {\cal L} \right\}$ (where ${\cal L} ( \cdot ) = [\hat{H} , \cdot ]$ is the overall system Liouvillian and $\text{Tr}_n\{\cdot\}$ is the partial trace over the nuclear (bath) Hilbert space), which is represented by a $N_e^2 \times N_e^2$ time-independent matrix. The second term is given in terms of the memory kernel ${\cal K}(\tau)$, which is represented by a $N_e^2 \times N_e^2$ time-dependent matrix. 
%Further details about the GQME approach, including an outline of the derivation of Eq.~\eqref{eq:mGQME} and the explicit form of the memory kernel, are provided in Sec.~\ref{sec:methods}.

The GQME formalism provides a general framework for deriving the exact EoM for any quantity of interest. 
The derivation begins with the Nakajima-Zwanzig equation \cite{nakajima58,zwanzig60b},
which describes the dynamics of a projected state ${\cal P}\hat{\rho}(t)$, where ${\cal P}$ is a projection superoperator and $\hat{\rho} (t)$ is the density operator of the overall system:
\begin{eqnarray}
\frac{d}{dt} \mathcal{P} \hat{\rho} (t) = -\frac{i}{\hbar} \mathcal{P} \mathcal{L} \mathcal{P} \hat{\rho} (t) -\frac{1}{\hbar^2} \int_0^t d \tau \mathcal{P} \mathcal{L} e^{-i \mathcal{QL} \tau/\hbar} \mathcal{Q} \mathcal{L} \mathcal{P} \hat{\rho} (t-\tau) \label{dPrhodt}
\\  \quad -\frac{i}{\hbar} \mathcal{P} \mathcal{L} e^{-i \mathcal{QL} t/\hbar} \mathcal{Q} \hat{\rho} (0).  \nonumber
\end{eqnarray}
Here, ${\cal L}$ is the overall system-bath Liouvillian and ${\cal Q} = \mathit{1} - {\cal P}$ is the complimentary projection superoperator to ${\cal P}$. Importantly, the only requirements are that ${\cal L}$ is Hermitian and $\cal{P}$ satisfies ${\cal P}^2 = \cal{P}$. Otherwise, there is complete flexibility in the choice of ${\cal L}$ and ${\cal P}$, with each choice leading to a different GQME for a different quantity of interest~\cite{mulvihill22}. 

Following Ref.~\citenum{mulvihill19a} we focus an overall system-bath Hamiltonian of the form of Eq.~(\ref{eq:genH}) and the following choice of projection operator which gives rise to the GQME for the system reduced density matrix, $\hat{\sigma} (t)$:
%for an overall system-bath Hamiltonian of the form of Eq.~(\ref{eq:genH}):
\begin{equation}
\mathcal{P} ( \hat{A}) = \hat{\rho}_n(0) \otimes \text{Tr}_n\{\hat{A}\}.
\label{eq:Proj_full}
\end{equation}
With this choice of $\mathcal{P}$, we have $\mathcal{Q}(\hat{\rho}_n(0))=0$. Plugging Eq.~(\ref{eq:Proj_full}) into Eq.~(\ref{dPrhodt}) and tracing over the nuclear (bath) Hilbert space leads to the GQME in Eq.~\eqref{eq:mGQME}.
The memory kernel in Eq.~\eqref{eq:mGQME} is given by
\begin{equation}
\mathcal{K} (\tau) = \frac{1}{\hbar^2} \text{Tr}_n \Big\{\mathcal{L}\,e^{-i\mathcal{Q} \mathcal{L}\tau/\hbar}\mathcal{Q} \mathcal{L}\hat{\rho}_n (0) \Big\},
\label{eq:K_full}
\end{equation}
and can be obtained by solving the following Volterra equation~\cite{mulvihill19a}:
\begin{align}
\mathcal{K} (\tau) &= i \dot{\mathcal{F}} (\tau) -\frac{1}{\hbar} \mathcal{F} (\tau) \langle \mathcal{L} \rangle_n^0 + i \int_0^\tau d\tau' \mathcal{F} (\tau - \tau') \mathcal{K} (\tau'). \label{volterra2}
\end{align}
Here, $\mathcal{F} (\tau)$ and $\dot{\mathcal{F}} (\tau)$ are the so-called projection-free inputs (PFIs), which are given by
\begin{equation}
\begin{split}
\mathcal{F} (\tau) &= \frac{1}{\hbar} \text{Tr}_n \Big\{ \mathcal{L} e^{-i \mathcal{L} \tau / \hbar} \hat{\rho}_n (0) \Big\},
\\ \dot{\mathcal{F}} (\tau) &= -\frac{i}{\hbar^2} \text{Tr}_n \Big\{ \mathcal{L} e^{-i \mathcal{L} \tau / \hbar} \mathcal{L} \hat{\rho}_n (0) \Big\}.
\end{split}
\label{proj_free_F}
\end{equation}
The memory kernels for the spin-boson model used in this paper were adopted from Ref.~\citenum{lyu2023tensor}, where they were obtained from quantum-mechanically exact PFIs calculated via the tensor-train thermo-field dynamics (TT-TFD) method.

%From Eq.~(\ref{eq:mGQME}), we can obtain a GQME for
The quantum open system's {\em non-unitary} time evolution superoperator, or propagator, ${\cal G} (t)$, is defined by:
\begin{equation}
\hat{\sigma} (t) = {\cal G} (t) \hat{\sigma} (0)~~. 
\label{eq:G_nonunitary}
\end{equation}
Substituting Eq.~(\ref{eq:G_nonunitary}) into Eq.~(\ref{eq:mGQME}) and noting that the GQME should be satisfied for an arbitrary choice of $\hat{\sigma} (0)$, it is straightforward to show that ${\cal G} (t)$ satisfies the same GQME as $\hat{\sigma} (t)$:
\begin{eqnarray}
\frac{d}{dt} {\cal G}(t) = -\frac{i}{\hbar} \langle \mathcal{L} \rangle_n^0 {\cal G}(t)- \int_0^t d\tau\ \mathcal{K}(\tau){\cal G}(t - \tau)~~.
\label{eq:mGQME-G}
\end{eqnarray} 
Thus, given the projected Liouvillian and memory kernels ($\langle \mathcal{L} \rangle_n^0$ and ${\cal K} (\tau)$, respectively), ${\cal G}(t)$ can be obtained by solving Eq.~(\ref{eq:mGQME-G}) numerically , which in this work was accomplished via a Runge–Kutta fourth-order (RK4) algorithm~\cite{mulvihill21a}. This superoperator, ${\cal G}(t)$, serves a role similar to that of the Kraus operators in the operator sum representation and can also be dilated to  a unitary form which can be implemented on a quantum computer. Importantly, while the Kraus operators are 
only known in closed form for
the Markovian Lindblad equation, the 
%time evolution superoperator 
non-unitatry propagator ${\cal G}(t)$ can always be obtained from the formally exact GQME (see Eq.~\eqref{eq:mGQME-G}).

%\subsection{TT-TFD method}
%\label{sec:methods}

%%%%%%%%%%%%%%%
\subsection{A GQME-based quantum algorithm for simulating open quantum system dynamics}
\label{subsec:quantum algorithm}

In this subsection, we describe the next step in the workflow outlined in Fig.~\ref{fig:workflow}, namely using the Sz.-Nagy’s unitary dilation procedure~\cite{levy2014dilation}
to convert the non-unitary quantum open system propagator
%time evolution superoperator 
${\cal G} (t)$ [see Eqs.~(\ref{eq:G_nonunitary}) and (\ref{eq:mGQME-G})] into a unitary propagator 
%time evolution superoperator 
in an extended Hilbert space. 
It should be noted that the Sz.-Nagy unitary dilation procedure is one 
%example of the general dilation 
out of several methods that can convert non-unitary operators into unitary operators
%, and there also exist other methods that achieve the same purpose such as the 
(e.g. block-encoding represents an alternative method~\cite{gilyen2019quantum,camps2022explicit}).

The Sz.-Nagy’s unitary dilation procedure starts out by calculating the operator norm of ${\cal G}(t)$ to determine if it is a \textit{contraction}. For ${\cal G}(t)$
to be a contraction, the operator norm of ${\cal G}(t)$ needs to be less than or equal to $1$, i.e., $||{\cal G}(t)||_O=\sup\frac{||{\cal G}(t)\bm{v}||}{||\bm{v}||}\leq 1$. 
% what is v in this equation? - EM
In the case where the original ${\cal G}(t)$ is {\em not} a contraction, we introduce a normalization factor $n_c=||{\cal G}(t)||_O$ in order to define a contraction form of ${\cal G}(t)$, namely ${\cal G}^{\prime}(t)={\cal G}(t)/n_c$. 

In the next step, we apply a $1$-dilation procedure to 
${\cal G}^{\prime}(t)$ to obtain a unitary ${\cal U}_{{\cal G}^{\prime}}(t)$ in an extended Hilbert space of double the dimension of the original system's Hilbert space:
\begin{equation}
\label{eq:1-dilation}
    {\cal U}_{{\cal G}^{\prime}}(t) = \begin{pmatrix}
        {\cal G}^{\prime}(t) & {\cal D}_{{\cal G}^{\prime \dagger}}(t) \\
        {\cal D}_{{\cal G}^{\prime}}(t) & -{\cal G}^{\prime \dagger}(t)
    \end{pmatrix}~~.
\end{equation}
Here, ${\cal D}_{{\cal G}^{\prime}}(t) =\sqrt{I-{\cal G}^{\prime \dagger}(t){\cal G}^{\prime}(t)}$ and ${\cal D}_{{\cal G}^{\prime \dagger}}=\sqrt{I-{\cal G}^{\prime}(t){\cal G}^{\prime \dagger}(t)}$, where ${\cal D}_{{\cal G}^{\prime}}(t)$ is the so-called defect superoperator of ${\cal G}^{\prime}(t)$. The 1-dilation procedure generates a unitary superoperator ${\cal U}_{{\cal G}^{\prime}}(t)$ that operates in the extended Hilbert space and replicates the effect of the contraction form of the original time evolution superoperator, ${\cal G}^{\prime}(t)$, when the input and output vectors are both 
%restricted to 
projected onto the original smaller Hilbert space. 

In the original system's Hilbert space, the system reduced density operator $\hat\sigma(t)$ %in \S~\ref{subsec: the spin-boson model} 
is represented by an $N_e \times N_e$ matrix:
\begin{equation}
   \hat \sigma(t) \doteq \left( \begin{matrix}
   {{\sigma }_{11}}(t) & \ldots  & {{\sigma }_{1N_e}}(t)  \\
   \vdots  & \ddots  & \vdots   \\
   {{\sigma }_{N_e1}}(t) & \cdots  & {{\sigma }_{N_eN_e}}(t)  \\
\end{matrix} \right).
\end{equation}
Alternatively, the same system reduced density operator can also be  represented by an $N_e^2$-dimensional vector in Liouville space:
\begin{equation}
    \hat\sigma(t) \doteq {{\left( {{\sigma }_{11}}(t),...,{{\sigma }_{1N_e}}(t),... \, ...,{{\sigma }_{N_e1}}(t),...,{{\sigma }_{N_eN_e}}(t) \right)}^{T}}~~.
\end{equation}
%\begin{equation}
%    \hat\sigma(t) \doteq {{\left( {{\sigma }_{11}}(t),...,{{\sigma }_{1N_e}}(t),{{\sigma }_{21}}(t),...,{{\sigma }_{2N_e}}(t),... \, ...,{{\sigma }_{N_e1}}(t),...,{{\sigma }_{N_eN_e}}(t) \right)}^{T}}~~.
%\end{equation}
Since the GQME formalism is given in terms of superoperators, it is convenient to work in Liouville space, which we will do from this point on.
%, In what follow we use the Liouville space form Because we are working with superoperators, we consider the electronic reduced density operator in Liouville space and therefore in vector form.
We also define the norm of the vector representing $\hat{\sigma} (t)$ in Liouville space as the 
%To adopt the form of the vector ${\sigma }(t)$ to that of an initial quantum input state, we calculate the 
Frobenius norm:
%of ${\sigma }(t)$ as 
$\left\| {{\sigma }(t)} \right\|_F=\sqrt{\sum\limits_{ij}{{{\left| {{\sigma }_{ij}} \right|}^{2}}}}$ and divide $\hat{\sigma }(t)$ by $\left\| {{\sigma }(t)} \right\|_F$ to normalize $\hat{\sigma }(t)$.~\cite{hu2020quantum} 

Given the dilated unitary operator ${\cal U}_{{\cal G}^{\prime}}(t)$ and the initial quantum input state ${\hat{\sigma} }(0)$, operating with the non-unitary ${\cal G}'(t)$ on $\hat \sigma(0)$ has now been converted into a unitary transformation as follows:
\begin{equation}
\label{eq:dilation process}
    {\cal G}^{\prime}(t)\hat{\sigma}(0)\xrightarrow{\text{unitary dilation}}{\cal U}_{{\cal G}^{\prime}}(t) \left( \hat\sigma(0)^{T},0,\cdots,0 \right)^{T}.
\end{equation}
The $0$s in the input vector on the R.H.S. are  added to match the dimension of the input vector with that of ${\cal U}_{{\cal G}^{\prime}}(t)$. The unitary process can then be simulated on a quantum circuit with unitary quantum gates. The  electronic populations, $\{ \sigma_{jj}(t) \equiv \langle j | \hat{\sigma} (t) | j \rangle | j=1,...,N_e\}$ can be retrieved by taking the square roots of  the probability of measuring each basis state $P_j(t)=|\sigma^{\prime}_{jj}(t)|^2$ and multiplying by the $n_c$ factor. 

Finally, we perform a complexity analysis of the quantum algorithm. Given that ${\cal G}(t)$ in its most general form is represented by a matrix of $N_e^4$ non-zero elements, the defect superoperators ${\cal D}_{{\cal G}^{\prime}}(t)$ as well as $-{\cal G}^{\dagger}(t)$ as shown in Eq.~\eqref{eq:1-dilation} all have $N_e^4$ non-zero elements. 
Generally speaking, the number of the two-level unitaries necessary to decompose a unitary gate is comparable to the number of non-zero elements in the lower-triangular part of the gate~\cite{Nielsen2011,Reck1994}. Therefore, the gate complexity to simulate this specific ${\cal U}_{{\cal G}^{\prime}}(t)$ is $O(N_e^4)$. If the two-level unitaries are further decomposed into $1$-qubit and $2$-qubit elementary gates commonly used to design conventional quantum circuits, they need to be transformed to the Gray code sequences and some multi-control gate sequences, adding another factor of complexity logarithmic in $N_e^2$, and the total complexity becomes $O(N_e^4\log^2 N_e^2)$~\cite{Nielsen2011}. This means that the maximum total complexity of a GQME-based simulation of a open quantum system dynamics 
is comparable to classical methods~\cite{hu2020quantum}. However, as demonstrated in previous simulations of certain dynamical models, our quantum algorithm can take advantage of the case when the  ${\cal G}(t)$ is a sparse matrix, and thus the gate complexity scaling for ${\cal G}(t)$  can be reduced to $O(\log^2 N_e^2)$ instead of $O(N_e^4)$~\cite{hu2020quantum,hu2022general}.

%%%%%%%%%%%%%%%%%%%%%%%%%%%%%%%%%%%%%%%%%%%%%%%%%%%%%
\section{Results}

\subsection{A demonstrative application to the spin-boson model }
\label{subsec: the spin-boson model}

In this subsection, we test the applicability of the quantum algorithm outlined in the previous sections on the spin-boson benchmark model. 
%The spin-boson model provides a general framework for modeling a system with 
%This model has the form of the Hamiltonian in Eq.~(\ref{eq:genH}), with two coupled electronic states which are also coupled to a bath of harmonic nuclear modes. 
This model and its derivatives have a wide range of applicability to chemical and physical systems, including electron, proton, energy, and charge transfer processes; polaron formation and dynamics in condensed phase environments; vibrational relaxation, impurity relaxation in solids, spin-lattice relaxation, and qubit decoherence~\cite{leggett87, breuer02, nitzan06, weiss12}. 
%For this reason, the spin-boson model has become an important benchmark model for the simulation of the dynamics of quantum open systems.
It should also be noted that quantum-mechanically exact memory kernels  for this model are available~\cite{shi03g,chatterjee19,lyu2023tensor}.
%\textcolor{red}{Please add reference to bib file.-- E. Geva}

The spin-boson Hamiltonian has the form of Eq.~\eqref{eq:genH} with $N_e=2$
and $\{ \hat{H}_j , \hat{V}_{jk} \}$ given by:
\begin{equation}
\begin{split}
\hat{H}_0 &\equiv \hat{H}_D = \epsilon + \sum_{k = 1}^{N_n} \frac{\hat{P}_k^2}{2} + \frac{1}{2}\omega_k^2\hat{R}_k^2 -c_k \hat{R}_k,
\\\hat{H}_1 &\equiv \hat{H}_A = -\epsilon + \sum_{k = 1}^{N_n} \frac{\hat{P}_k^2}{2} + \frac{1}{2}\omega_k^2\hat{R}_k^2 +c_k \hat{R}_k,
\\ 
\hat{V}_{01} &\equiv \hat{V}_{DA} = \hat{V}_{10} \equiv \hat{V}_{AD} = \Gamma.
\end{split} 
\label{eq:SBham}
\end{equation}  
Here, the two electronic states are designated as the donor and acceptor ($| D \rangle$ and $| A \rangle$, respectively), $2 \epsilon$ is the shift in equilibrium energy between the donor ($D$) and acceptor ($A$) states, and $\Gamma$ is a positive constant describing the electronic coupling between the donor and acceptor states. Since $\Gamma$ is a constant, this system is assumed to satisfy the Condon approximation.
%\textcolor{red}{The reason why is not clear here.--M.Soley}

The results shown below were obtained for the case where the nuclear modes' frequencies and coupling coefficients $\{\omega_k , c_k\}$ are sampled from an Ohmic spectral density with exponential cutoff:
%{It would be beneficial to explain why we sample from an Ohmic distribution.--M.soley}
\begin{equation}
  J (\omega)  = \frac{\pi}{2} \sum_{k=1}^{N_n} \frac{c_k^2}{\omega_k} \delta(\omega-\omega_k) ~  \stackrel{\raisebox{1pt} {\text{\footnotesize$N_n \rightarrow \infty$}}}{\xrightarrow{\hspace*{0.75cm}}} ~ \frac{\pi\hbar}{2}
 \xi \omega e^{-\omega/\omega_c} ~~.
 \label{ohmic}
\end{equation}
Here, $\xi$ is the Kondo parameter and $\omega_c$ is the cutoff frequency. The reader is referred to Appendix C of Ref.~\citenum{mulvihill19a} for a description of the  procedure used to obtain a discrete set of $N_n$ nuclear mode frequencies $\{\omega_k\}$ and coupling coefficients $\{c_k\}$ from the spectral density in Eq.~(\ref{ohmic}).

\begin{table}
\centering
\caption{Spin-boson model and simulation parameters. 
%The units of the parameters are given in terms of the electronic coupling $\Gamma$.
}
\def\arraystretch{1.125}
\resizebox{1.\columnwidth}{!}{
\begin{tabular}{|c||c|c|c|c|c||c|c|c|}
  \hline
& \multicolumn{5}{c||}{Model Parameters} & \multicolumn{3}{c|}{Numerical Parameters}
\\ \hline Model $\#$ & $\epsilon$ & $\Gamma$ & $\beta$ & $\xi$ & $\omega_c$  &\ $\omega_{\text{max}}$\ \ & \ \ $N_n$\ \ & $\Delta t$  
  \\ \hline
  1  & 1.0\,$\Gamma$  & 1.0  & 5.0\,$\Gamma^{-1}$  & 0.1  & 1.0\,$\Gamma$  & 5\,$\Gamma$   & 60 & 1.50083 $\times 10^{-3}\,\Gamma^{-1}$
  \\ \hline
  2  & 1.0\,$\Gamma$  & 1.0  & 5.0\,$\Gamma^{-1}$  & 0.1  & 2.0\,$\Gamma$  & 10\,$\Gamma$  & 60 & 1.50083 $\times 10^{-3}\,\Gamma^{-1}$
  \\ \hline
  %3  & 1.0  & 1.0  & 5.0  & 0.1  & 7.5  & 36  & 60 & 1.50083 $\times 10^{-3}$
  %\\ \hline
  3  & 1.0\,$\Gamma$  & 1.0  & 5.0\,$\Gamma^{-1}$  & 0.4  & 2.0\,$\Gamma$  & 10\,$\Gamma$  & 60 & 1.50083 $\times 10^{-3}\,\Gamma^{-1}$
  \\ \hline
  4  & 0.0\,$\Gamma$  & 1.0  & 5.0\,$\Gamma^{-1}$  & 0.2  & 2.5\,$\Gamma$  & 12\,$\Gamma$  & 60 & 4.50249 $\times 10^{-3}\,\Gamma^{-1}$
  \\ \hline
\end{tabular} 
}
\label{tab:parameters}
\end{table}

The initial state is assumed to be of the form of Eq.~(\ref{init_stat}), with the initial electronic (system) reduced density operator given by 
\begin{equation}
\hat{\sigma} (0) = | D \rangle \langle D |~~
\label{eq:sigma0}
\end{equation}
and the initial nuclear (bath) reduced density operator given by
\begin{equation}
  \hat{\rho}_n (0) 
%  = \frac{e^{-\beta \hat{H}_B}}
%  {\text{Tr}_B \Big\{ e^{-\beta \hat{H}_B } \Big\}}
  =\frac{e^{-\beta (\hat{H}_D + \hat{H}_A)/2}}{\text{Tr}_n \Big\{ e^{-\beta(\hat{H}_D + \hat{H}_A)/2} \Big\}}.
  \label{init-nuc}
\end{equation}

Calculations were carried out for four different sets of parameter values
%, with units scaled to the electronic coupling $\Gamma$ 
(see Table \ref{tab:parameters}). Models 1 and 2 correspond to systems with an energy bias between the donor and acceptor states ($\epsilon \neq 0$) and differ in their cutoff frequencies, with model 2 having a higher cutoff frequency. Model 3 corresponds to a biased system with the same parameters as model 2 except for a larger Kondo parameter. Model 4 corresponds to a symmetric system with zero energy bias between the donor and acceptor states ($\epsilon = 0$). The results reported in this paper were obtained with a time step of $\Delta t = 1.50083 \times 10^{-3}\,\Gamma^{-1}$ for models 1-3 and a time step of $\Delta t = 4.50249 \times 10^{-3}\,\Gamma^{-1}$ for model 4. 
%\textcolor{red}{It would be beneficial to describe the physical meanings of the models.--M.Soley}

Starting with the quantum-mechanically exact memory kernels (adopted from Ref.~\citenum{lyu2023tensor}),  the time evolution superoperator for the electronic reduced density matrix ${\cal G}(\tau)$ was generated for the four models given in Table \ref{tab:parameters} by solving the corresponding GQME, Eq.~(\ref{eq:mGQME-G}). 

\begin{figure}[h!]
\centering
    \subfloat[Model 1]{
	\begin{minipage}[t]{0.45\linewidth}
	   \centering
	   \includegraphics[width=1\textwidth]{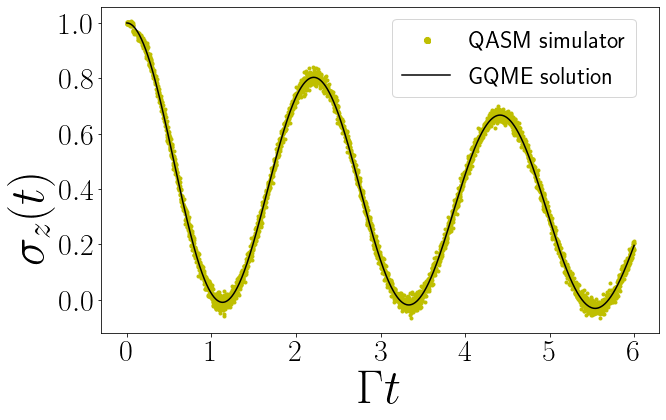}
	\end{minipage}}
 	\subfloat[Model 2]{
	\begin{minipage}[t]{0.45\linewidth}
	   \centering
	   \includegraphics[width=1\textwidth]{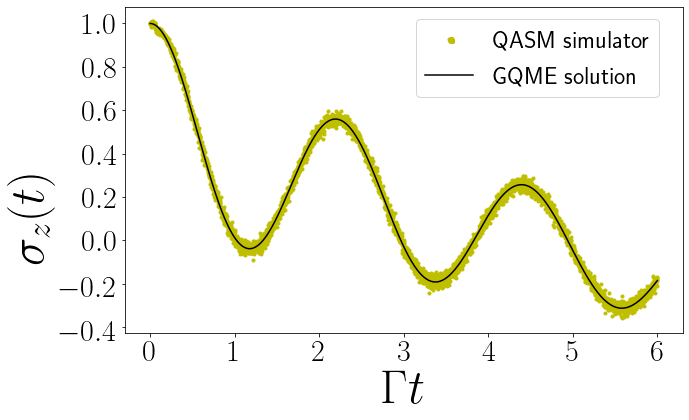}
	\end{minipage}}

  \subfloat[Model 3]{
	\begin{minipage}[b]{0.45\linewidth}
	   \centering
	   \includegraphics[width=1\textwidth]{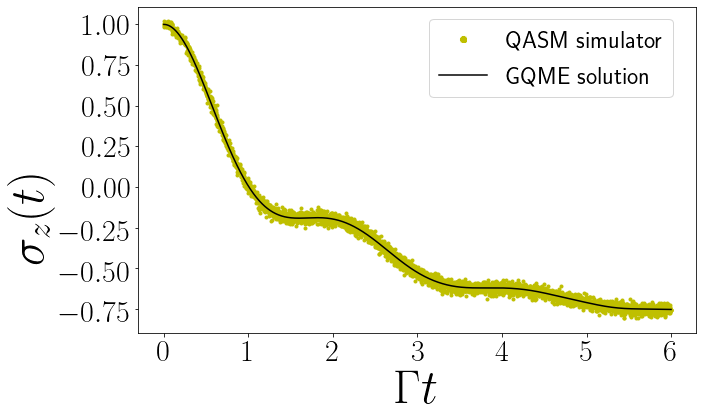}
	\end{minipage}}
    \subfloat[Model 4]{
	\begin{minipage}[b]{0.45\linewidth}
	   \centering
	   \includegraphics[width=1\textwidth]{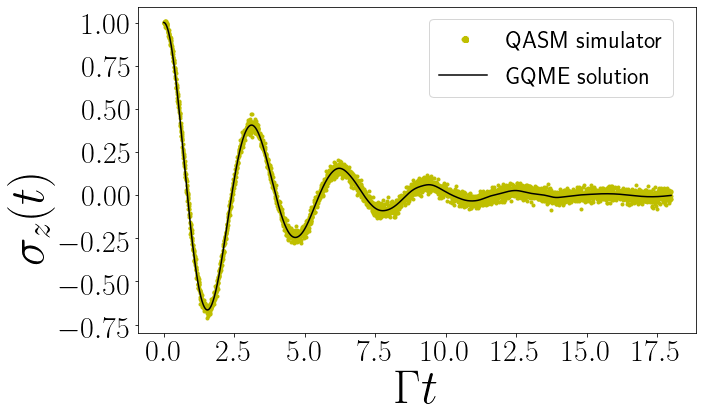}
	\end{minipage}}
    
\caption{The spin-boson model simulated by the GQME-based quantum algorithm as implemented on the IBM QASM quantum simulator, showing the electronic population difference between the donor state and acceptor state $\sigma_z(t) =\sigma_{DD}(t) -\sigma_{AA}(t)$ as a function of time for (a) model 1, (b) model 2, (c) model 3, and (d) model 4 as given in Table \ref{tab:parameters}, with units scaled to the electronic coupling, $\Gamma$.
%with units scaled to the electronic coupling $\Gamma$.
Each figure shows the comparison between the GQME-based exact results represented by the black curves and the QASM-based results represented by the yellow dots. The time step for both  the exact and simulated results is $\Delta t=1.50083 \times 10^{-3}\Gamma^{-1}$ for models 1-3 and $\Delta t = 4.50249 \times 10^{-3}\Gamma^{-1}$ for model 4. Each model is simulated for $4000$ time steps. The number of projection measurements applied by the QASM simulator to obtain a single time step is $2000$ shots.}
	\label{fig:QASM_simulation}
\end{figure}

\begin{figure}[h!]
\centering
	\subfloat[Model 1]{
	\begin{minipage}[t]{0.45\linewidth}
	   \centering
	   \includegraphics[width=1\textwidth]{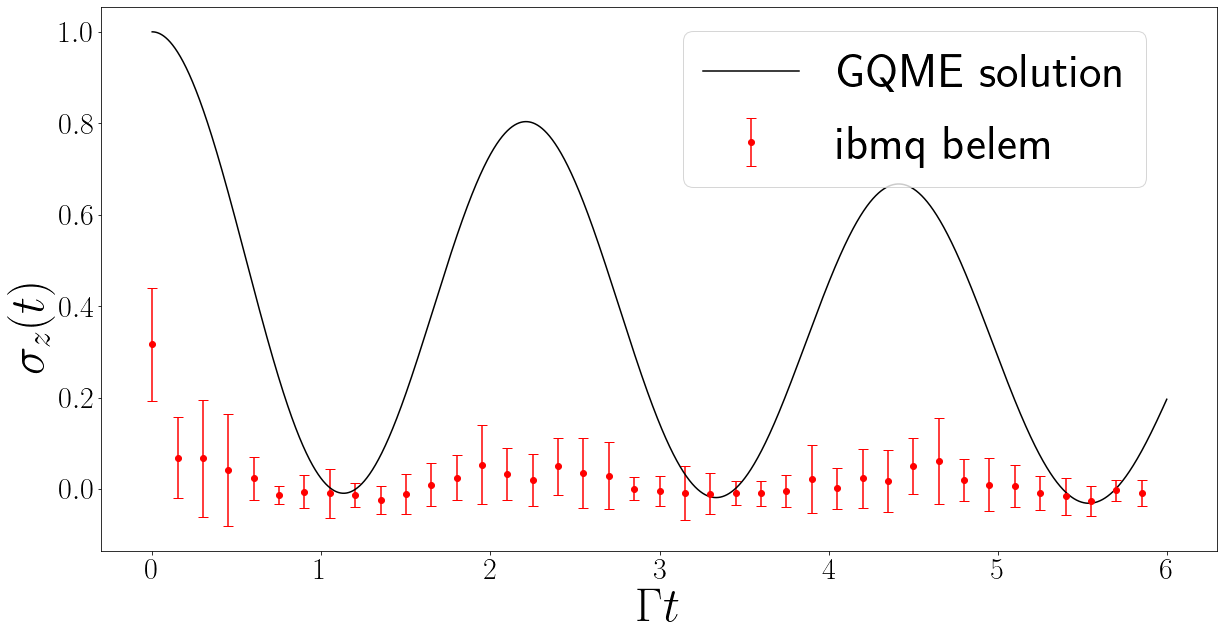}
	\end{minipage}}
 \subfloat[Model 2]{
	\begin{minipage}[t]{0.45\linewidth}
	   \centering
	   \includegraphics[width=1\textwidth]{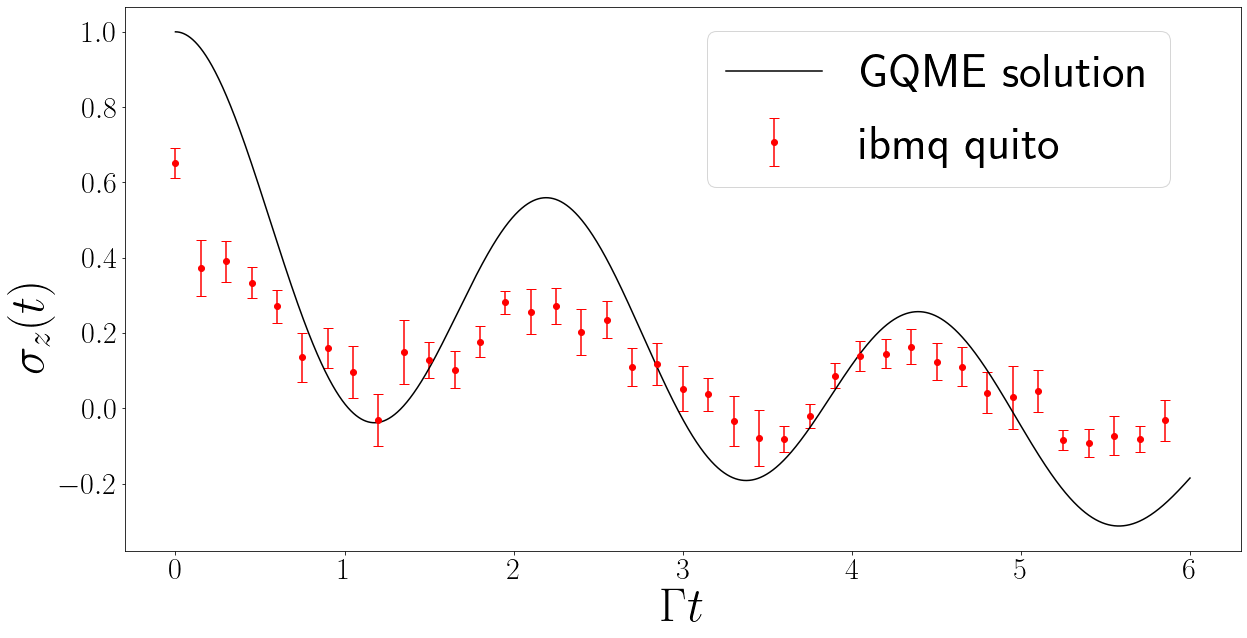}
	\end{minipage}}
 
 	\subfloat[Model 3]{
	\begin{minipage}[t]{0.45\linewidth}
	   \centering
	   \includegraphics[width=1\textwidth]{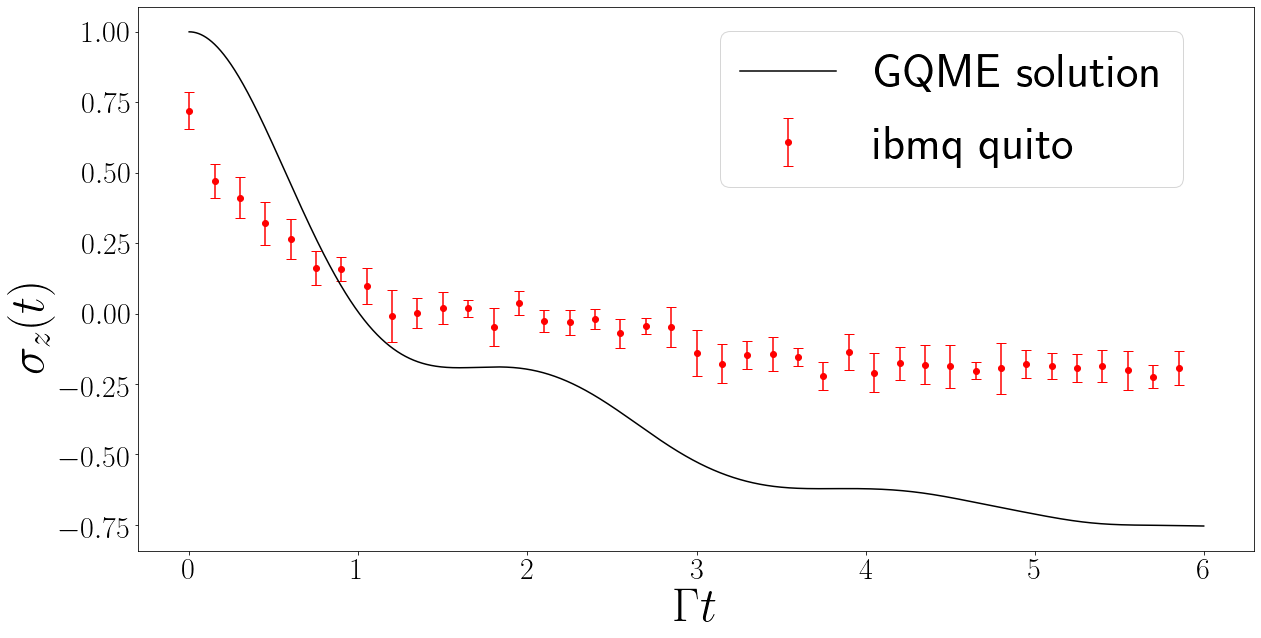}
	\end{minipage}}
 \subfloat[Model 4]{
	\begin{minipage}[t]{0.45\linewidth}
	   \centering
	   \includegraphics[width=1\textwidth]{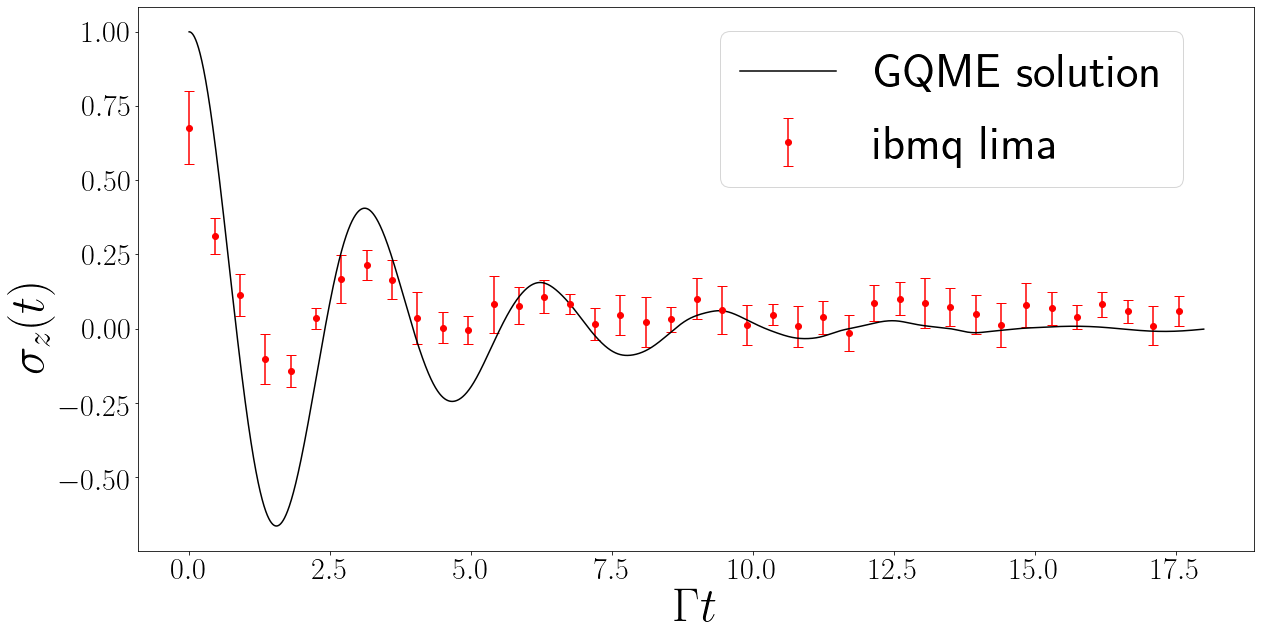}
	\end{minipage}}
\caption{The spin-boson model simulated by the GQME quantum algorithm as implemented on the IBM quantum computers \textbf{ibmq belem}, \textbf{ibmq quito} and \textbf{ibmq lima}, showing the electronic population difference between the donor state and acceptor state $\sigma_z(t) =\sigma_{DD}(t) -\sigma_{AA}(t)$ as a function of time for (a) model 1 , (b) model 2 (c) model 3 and (d) model 4 as given in Table \ref{tab:parameters}, with units scaled to the electronic coupling, $\Gamma$. Each figure shows the comparison between the GQME-based exact results represented by the black curves and quantum-computer-based results represented by the red dots with error bars. The time step for the real machine simulation is $\Delta t=0.150083\, \Gamma^{-1}$ for model 1,2 and 3 and $\Delta t=0.450249\,\Gamma^{-1}$ for model 4. The experiments of both models take 40 evenly-spaced time steps out of the 4000 time steps used in the QASM simulator runs and the error bars represent the standard derivations of the $10$ separate runs on the \textbf{ibmq belem}, \textbf{ibmq quito} and \textbf{ibmq lima} for models 1 to 4. The number of projection measurements applied by all the devices to obtain a single time step is 2000 shots.}
    \label{fig:GQME_TFDTT_model_real}
\end{figure}

The GQME-based quantum algorithm for simulating the electronic dynamics within the spin-boson model was implemented on the IBM quantum platforms via the  Qiskit package~\cite{Qiskit}.
The quantum implementation involved the translation of ${\cal G}^{\prime}(t)$ into ${\cal U}_{{\cal G}^{\prime}}(t)$ at each time step, followed by the construction of a quantum circuit based on ${\cal U}_{{\cal G}^{\prime}}(t)$, and lastly the use of the quantum circuit to simulate the time evolution of the reduced electronic density matrix. 
%We first calculate the operator norms of all the ${\cal G}(t)$ matrices of the same model and set the $n_c$ factor to be the largest operator norm among them. Then we obtain the contraction ${\cal G}^{\prime}(t)={\cal G}(t)/n_c$. 
To build the circuit,
%implementing the ${\cal G}^{\prime}(t)$ 
we dilated the $4\times 4$ ${\cal G}^{\prime}(t)$ into a unitary $8\times 8$ ${\cal U}_{{\cal G}^{\prime}}(t)$ by using a $1$-dilation procedure [see Eq.~~\eqref{eq:1-dilation}]. The unitary ${\cal U}_{{\cal G}^{\prime}}(t)$ was then transpiled into a $3$-qubit quantum circuit composed of three elementary quantum gates: $R_Z$, $\sqrt{X}$, and $CX$. Examples of ${\cal U}_{{\cal G}^{\prime}}(t)$ and details of the elementary quantum gates and circuits are given in the supplementary information (SI).
The initial electronic state %, which is the flattened $\hat{\sigma} (0)$, 
is set to ${{\left( 1,0,0,0,0,0,0,0 \right)}^{T}}$, where the last four $0$s are the extra dimensions from the dilation procedure. The QASM simulator and the real quantum devices initialize the input state ${{\left( 1,0,0,0,0,0,0,0 \right)}^{T}}$ and apply the unitary operation ${\cal U}_{{\cal G}^{\prime}}(t)$ to the input state followed by projection measurements to retrieve the probability distribution of all the $8$ basis states. Each circuit runs 2000 shots and the resulting probabilities $P_{000}(t)$ of measuring the state $|000\rangle$  and $P_{011}(t)$ of measuring $|011\rangle$  correspond to the diagonal elements of the modified density matrix $|\sigma^{\prime}_{00}(t)|^2$ and $|\sigma^{\prime}_{11}(t)|^2$. The populations of the donor state, $\sigma_{00}(t)$, and acceptor state, $\sigma_{11}(t)$, are retrieved as follows:
\begin{equation}
    \sigma_{00}(t) = \sqrt{P_{000}(t)}\times n_c \;\mathrm{and}\;
\sigma_{11}(t) = \sqrt{P_{011}(t)}\times n_c~~.
\end{equation}
In what follows, we report results in terms of the difference between the donor and acceptor populations, $\sigma_z(t) =\sigma_{00}(t) -\sigma_{11}(t)$. 

%The final result in Fig.~\ref{fig:QASM_simulation} is illustrated by the difference between the donor and acceptor states as $\sigma_z(t) =\sigma_{00}(t) -\sigma_{11}(t)$. 
The comparison between the exact results obtained by solving the GQME on a classical computer and results obtained by  performing the quantum algorithm on the QASM simulator is shown in Fig.~\ref{fig:QASM_simulation}.  The QASM simulator results are in excellent agreement with the exact results for all four models under consideration. The small amplitude oscillations of the QASM-based results around the exact results can be traced back to the inherent uncertainty associated with projection measurements. These results validate the GQME-based quantum algorithm and demonstrate its ability to reproduce results obtained via the GQME-based classical algorithm. 
%demonstrate the ability of the quantum algorithm  to simulate  general open quantum system dynamics described by the GQME.

To test the performance of the quantum algorithm on real quantum devices, we also performed the simulations on the quantum computers provided by IBM Quantum (IBM Q). The simulations were performed for models 1 to 4 on \textbf{ibmq quito}, \textbf{ibmq belem} and \textbf{ibmq lima}. All devices are equipped with $5$ qubits that have the same qubit connectivity and use  IBM's \textbf{Falcon r4T} processor with the same architecture.
In each simulation of a given model, three qubits were used and 10 repeated experiments were performed. In a single experiment, 40 time steps are chosen at an equal spacing out of the 4000 time steps used in the QASM simulations, i.e., the time step in each experiment is $100$ times greater than the time step used in the QASM simulations as listed in Table \ref{tab:parameters}. 
The average $CX$ gate error and readout error are $(1.191\times 10^{-2},5.194\times 10^{-2})$ for the \textbf{ibmq quito}, $(1.160\times 10^{-2},2.590\times 10^{-2})$ for the \textbf{ibmq belem} and $(1.032\times 10^{-2},2.834\times 10^{-2})$ for the \textbf{ibmq lima} as of the time of the experiments.
% EM: this sentence is unclear; is it (gate error, readout error) in the parentheses, with the first set corresponding to quito and the second to lima? If so, the ", respectively," should come after "lima" 
The quantum circuits are the same in both the QASM simulations and the real machine simulations. The transpiled quantum gate counts for each of the ${\cal U}_{{\cal G}^{\prime}}(t)$ superoperators are $153$ $R_Z$ gates, $98$ $\sqrt{X}$ gates, and $41$ $CX$ gates. The transpiling process is done internally by the Qiskit package and examples of the quantum circuits can be found in the SI. 

The comparison between the GQME-generated exact results and real machine simulations is given in Fig.~\ref{fig:GQME_TFDTT_model_real}. In the figure, the red dots are the average of the 10 experiments and the error bars  represent standard derivations of the 10 experiments. While the results obtained on the IBM Q quantum computers reproduce some of the trends exhibited by the exact results, the agreement is qualitative at best. The lack of quantitative agreement can be traced back to the rather extensive circuit depth, which makes the calculation susceptible to noise. %and errors. 
In the next section, we propose a way to lower the circuit depth and enhance the accuracy of the calculation on the IBM Q quantum computers by using reduced-dimensionality GQMEs.  

%which demonstrate that the GQME-based quantum algorithm can capture the essential features of the dynamics even with the noise and errors within real quantum devices.

\subsection{Reduced-dimensionality GQME-based 
%non-unitary 
propagators 
%for a subset of the open quantum system's reduced density matrix elements
}
\label{subsec:red_dim_gqme}

Since the quantum algorithm on the QASM simulator was able to accurately reproduce the exact results, as shown in Fig.~\ref{fig:QASM_simulation}, we attribute the lack of quantitative agreement between the exact results and the results obtained via the IBM Q quantum computers, as seen in Fig.~\ref{fig:GQME_TFDTT_model_real}, to noise within the real quantum devices. If so, reducing the circuit depth would improve the accuracy. 
%One way to verify this would be to develop a way to simulate the system with a shallower circuit. 
In this subsection, we validate this hypothesis by 
%develop a method for decreasing the circuit depth by %for the spin-boson model
reducing the dimensionality of the non-unitary propagator ${\cal G}(t)$, and thereby lowering the circuit depth to levels that allow for an accurate calculation on the NISQ quantum computers. 
%and therefore also the unitary time evolution superoperator ${\cal U}_{\cal G'}(t)$.

To this end, we take inspiration from reduced-dimensionality GQMEs, which correspond to EoMs for subsets of the open quantum system's reduced density matrix elements, rather than the full reduced density matrix.\cite{mulvihill22, lyu2023tensor}
%By focusing on a subset of the electronic DOF rather than all of them, the dimensionality of the GQME approach was reduced, including for the memory kernel. 
For example, for the spin-boson model described in Sec. \ref{subsec: the spin-boson model}, the memory kernel in the GQME for the full reduced density matrix, $\hat\sigma(t)$, is a $4\times 4$ matrix, while the memory kernel in the GQME for only the two populations (the diagonal elements of the reduced density matrix, $\sigma_{00}(t)$ and $\sigma_{11}(t)$) is a $2\times 2$ matrix. \cite{mulvihill22, lyu2023tensor} 
%The success of these reduced-dimensionality GQMEs provided the inspiration to develop a way to reduce the dimensionality of the quantum algorithm in this article.
Below, we demonstrate how one can take advantage of this reduced dimensionality to lower the circuit depth and thereby improve the accuracy of the simulation on quantum machines. 

For the spin-boson model under consideration in this paper, the electronic populations can be propagated using only the four corner elements of ${\cal G}(t)$, i.e.,
\begin{equation}
\left(\begin{array}{c} \sigma_{11}(t) \\ \sigma_{22}(t) \end{array}\right)
= \left(\begin{array}{cc} {\cal G}_{11,11}(t) & {\cal G}_{11,22}(t) \\ {\cal G}_{22,11}(t) & {\cal G}_{22,22}(t) \end{array}\right)
\left(\begin{array}{c} \sigma_{11}(0) \\ \sigma_{22}(0) \end{array}\right).
\label{G_pop}
\end{equation}
It should be noted that this equality only holds when 
%${\cal G}(t)$ is obtained with an exact input method and 
the initial electionic state is of the form %a linear combination of only the populations, i.e., 
$\hat\sigma(0) = \sum_{j = 1}^{N_e} \sigma_{jj}(0) |j\rangle\langle j|$, which is consistent with the initial state under consideration in this paper (see Eq. \eqref{eq:sigma0}). %exact TT-TFD input method and the spin-boson model explored in this paper. 
It should also be noted that Eq. \eqref{G_pop} is still exact, in the sense that the time evolution of $\sigma_{11}(t)$ and $\sigma_{00}(t)$ as described by the equation is exactly the same time evolution as described by Eq. \eqref{eq:G_nonunitary}. Thus, the only price one pays for the reduced dimensionality is the loss of the ability to simulate the dynamics of the off-diagonal matrix elements
$\sigma_{10}(t)$ and $\sigma_{01}(t)$. However, given that the primary goal is often to simulate the dynamics of electronic energy/charge transfer,  the populations of the corresponding electronic states is all that one needs. Finally, it is worth noting that our specific way of choosing the subset of the density operator does not indicate there is no coupling between the elements. In fact, such coupling can be captured exactly by the memory kernel and the effective Liouvillian of any open quantum system with the GQME.

%Similar to the procedure for the propagator for the full density matrix, ${\cal G}(t)$, %for all the electronic DOF, 
The $2 \times 2$  propagator in Eq. \eqref{G_pop}, which we will refer to as ${\cal G}^{\text{pop}}(t)$, can be dilated following a procedure similar to that we used to dilate the $4 \times 4$ propagator for the full density matrix, ${\cal G}(t)$. 
More specifically, ${\cal G}^{\text{pop}}(t)$
can be divided by a normalization factor $n_c^{\text{pop}} = ||{\cal G}^{\text{pop}}(t)||_O$ to obtain its contraction form ${\cal G}^{\text{pop}\prime}(t) = {\cal G}^{\text{pop}}(t)/n_c^{\text{pop}}$. 
Applying a 1-dilation procedure to ${\cal G}^{\text{pop}\prime}(t)$, similar to that 
in Eq.~\eqref{eq:1-dilation}, 
%we apply the 1-dilation procedure ${\cal G}^{\text{pop}\prime}(t)$ to 
then leads to the following unitary propagator:
%time evolution superoperator for the electronic populations ${\cal U}_{{\cal G}^{\text{pop}\prime}}(t)$:
\begin{equation}
    {\cal U}_{{\cal G}^{\text{pop}\prime}}(t) = \begin{pmatrix}
        {\cal G}^{\text{pop}\prime}(t) & {\cal D}_{{\cal G}^{\text{pop}\prime \dagger}}(t) \\
        {\cal D}_{{\cal G}^{\text{pop}\prime}}(t) & -{\cal G}^{\text{pop}\prime \dagger}(t)
    \end{pmatrix}.
    \label{eq:red_dim_U}
\end{equation}
Notably, for the spin-boson model, while ${\cal U}_{{\cal G}^{\prime}}(t)$ is an $8\times 8$ time-dependent matrix, ${\cal U}_{{\cal G}^{\text{pop}\prime}}(t)$ is a $4\times 4$ time-dependent matrix.

\begin{figure}[h!]
\centering
	\subfloat[Model 1]{
	\begin{minipage}[t]{0.45\linewidth}
	   \centering
	   \includegraphics[width=1\textwidth]{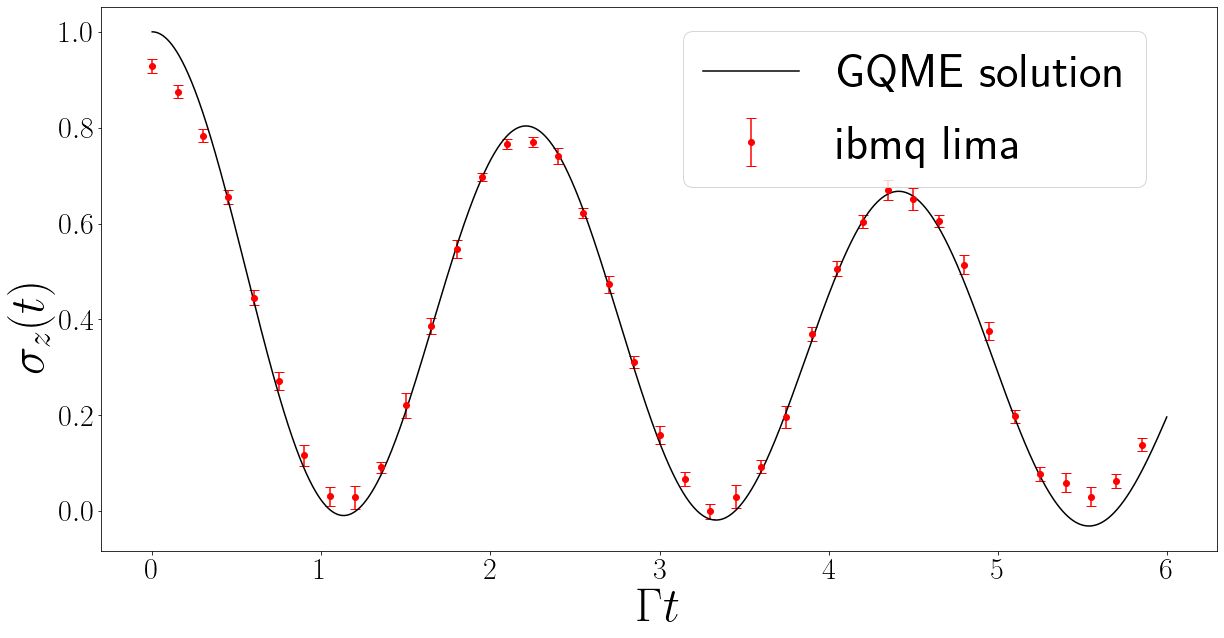}
	\end{minipage}}
 \subfloat[Model 2]{
	\begin{minipage}[t]{0.45\linewidth}
	   \centering
	   \includegraphics[width=1\textwidth]{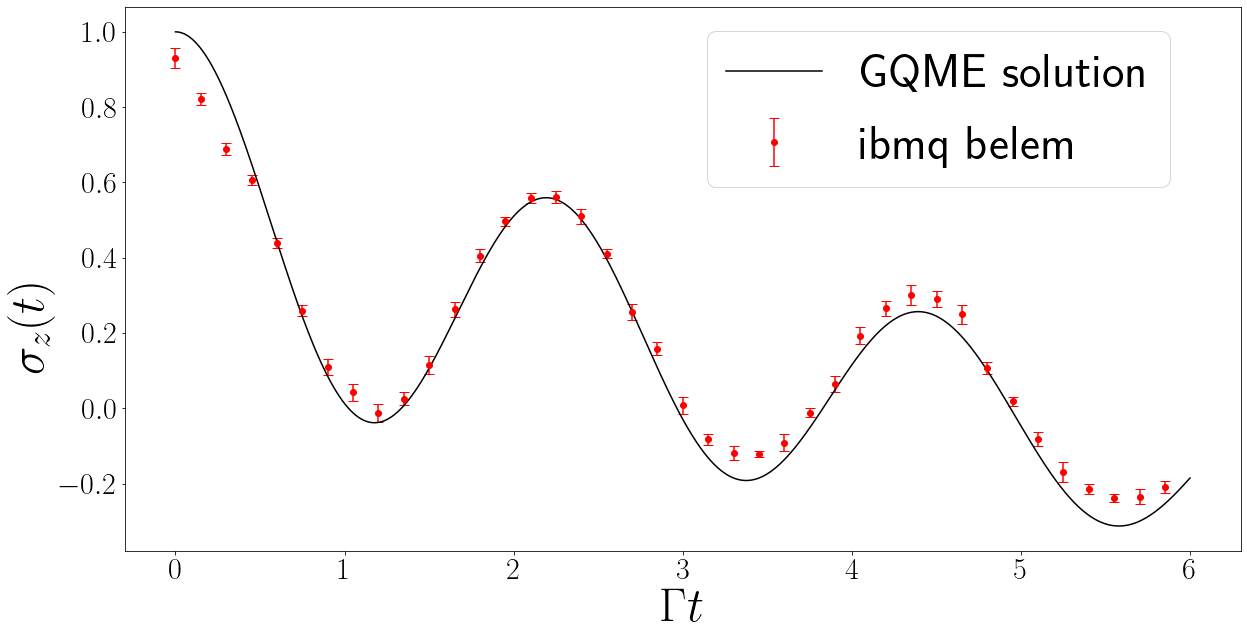}
	\end{minipage}}
 
 	\subfloat[Model 3]{
	\begin{minipage}[t]{0.45\linewidth}
	   \centering
	   \includegraphics[width=1\textwidth]{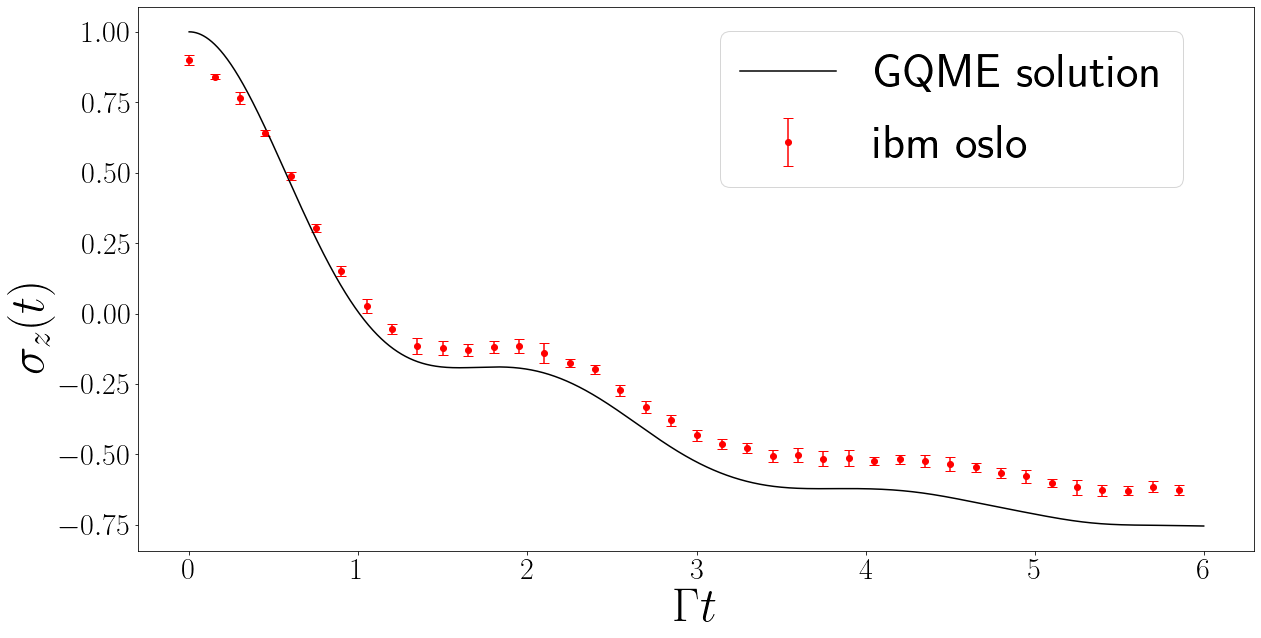}
	\end{minipage}}
 \subfloat[Model 4]{
	\begin{minipage}[t]{0.45\linewidth}
	   \centering
	   \includegraphics[width=1\textwidth]{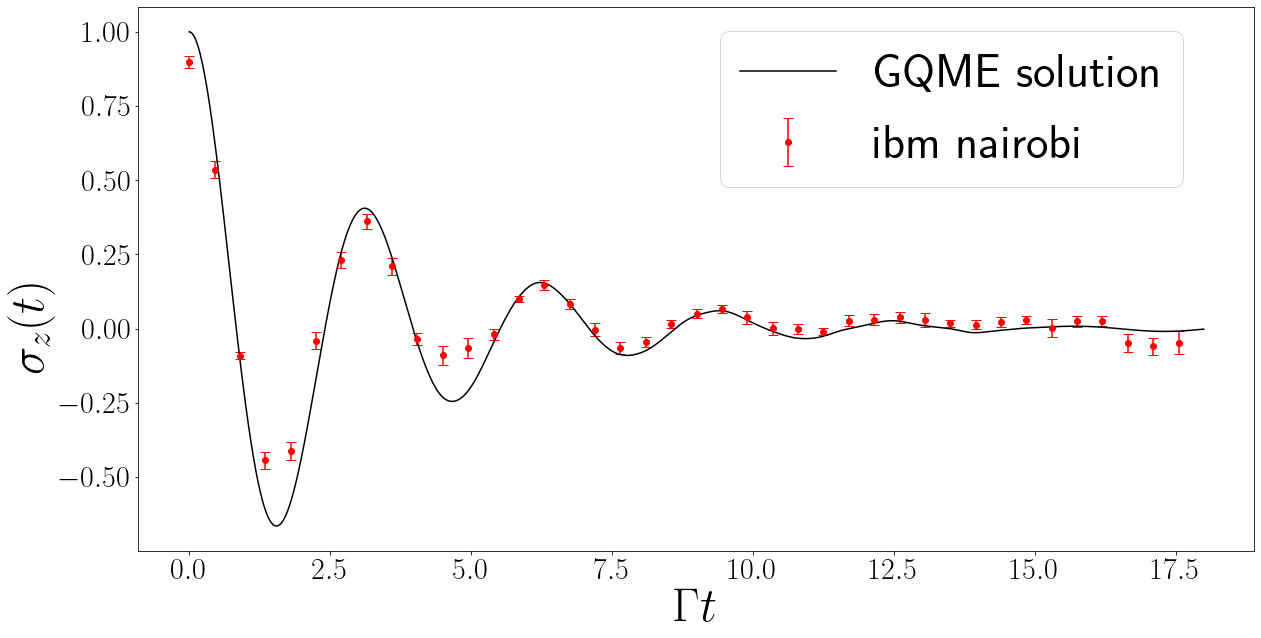}
	\end{minipage}}
\caption{
A comparison between the exact results for the spin-boson model and results obtained by performing the quantum algorithm based on Eq. \eqref{eq:red_dim_U} on the IBM Q quantum machines. 
The electronic population difference between the donor state and acceptor state, $\sigma_z(t) =\sigma_{DD}(t) -\sigma_{AA}(t)$, is plotted as a function of time for (a) model 1 , (b) model 2 (c) model 3 and (d) model 4 as given in Table \ref{tab:parameters}, with units scaled to the electronic coupling, $\Gamma$.
Each panel shows the comparison between the exact results  represented by the black curves and the population-only-GQME-based quantum-computer-simulated results represented by the red dots with error bars. The time step for the real machine simulation is $\Delta t=0.150083\, \Gamma^{-1}$ for model 1,2 and 3 and $\Delta t=0.450249\,\Gamma^{-1}$ for model 4. The experiments of both models take 40 evenly-spaced time steps out of the 4000 time steps used in the QASM simulator runs and the error bars represent the standard derivations of the $10$ separate runs on the \textbf{ibmq lima}, \textbf{ibmq belem}, \textbf{ibm oslo} and \textbf{ibm nairobi} for models 1-4, respectively. The number of projection measurements applied by all the devices to simulate a single time step is 2000 shots. 
}
\label{fig:reduced}
\end{figure}

A comparison between the exact results and results obtained by performing the quantum algorithm based on Eq. \eqref{eq:red_dim_U} on IBM Q quantum machines is shown in Fig. \ref{fig:reduced}.
%We simulated the same spin-boson models with the reduced-dimensionality GQMEs on the quantum computers provided by IBM Quantum (IBM Q). 
The results shown were obtained for models 1-4 on \textbf{ibmq belem}, \textbf{ibmq lima}, \textbf{ibm oslo} and \textbf{ibm nairobi}, respectively.
%and the comparison between the GQME-generated exact results and real machine simulations is given in Fig.\ref{fig:reduced}. 
Here, \textbf{ibm oslo} and \textbf{ibm nairobi} are each equipped with $7$ qubits of the same qubit connectivity and both use IBM's \textbf{Falcon r5.11H} processor. The average $CX$ gate error and readout error are $(1.038\times 10^{-2},2.28 0\times 10^{-2})$ for the \textbf{ibm nairobi} and $(8.537\times 10^{-3},2.310\times 10^{-2})$ for the \textbf{ibm oslo} as of the time of the experiments.
%In each simulation of a given model, three qubits were used and 10 repeated experiments were performed. In a single experiment, 40 time steps are chosen at an equal spacing out of the 4000 time steps used in the QASM simulations, i.e., the time step in each experiment is $100$ times greater than the time step used in the QASM simulations as listed in Table \ref{tab:parameters}. 
The new simulations use the same time steps, experiment shots and follow the same procedures as that used to obtain the results in Fig.~\ref{fig:GQME_TFDTT_model_real}.  
The quantum circuits are re-transpiled to implement the reduced-dimensionality GQME-based quantum algorithm where only two qubits are used. The transpiled quantum gate counts for each of the ${\cal U}_{{\cal G}^{\text{pop}\prime}}(t)$ superoperators are $17$ $R_Z$ gates, $12$ $\sqrt{X}$ gates, and $2$ $CX$ gates. The transpiling processes are done internally by the Qiskit package.

The results in Fig.~\ref{fig:reduced} confirm that the lack of quantitative agreement seen in Fig.~\ref{fig:GQME_TFDTT_model_real} 
%is not due to the weakness of the quantum algorithm, but 
can be attributed to noise on the real quantum devices. More specifically, significantly more accurate results are obtained when the populations-only reduced dimensionality GQME-based propagators are used, which can be traced back to their ability to give rise to shallower quantum circuits. 
Thus, reduced dimensionality 
%will potentially allow allows us 
make it possible to accurateley simulate  
%more complex and application-relevant 
the open quantum system dynamics on NISQ quantum computers. 

%One example of a follow-up study we are planning is to apply this method to simulate the exciton transfer dynamics of  the Fenna-Matthews-Olson complex  that is important in the photosynthesis process of bacteria. 

\section{Concluding Remarks}

The GQME-based quantum algorithm proposed herein substantially expands the range of open quantum systems that can be simulated on a quantum computer. In this paper, we demonstrated the applicability and versatility of the algorithm  by using it to simulate the dynamics of electronic populations within the benchmark spin-boson model on the IBM QASM quantum simulator and IBM quantum computers.

The results obtained via the noise-free QASM simulator were found to be highly accurate, with the only errors inherently associated with the quantum projection measurements and giving rise to very slight deviations from the exact results. 
However, while the implementation of the algorithm on the NISQ  IBM Q quantum computers was found to 
reproduce some of the trends exhibited by the exact results, the agreement was
qualitative at best. 
This lack of quantitative agreement was traced back to the rather extensive circuit depth, which made the calculation susceptible to noise. This issue was confirmed and fixed by implementing a populations-only reduced-dimensionality version of the quantum algorithm, which significantly shortened the circuit depth and as a result gave rise to quantitatively accurate results. 
%while results based on 
%the GQME for the full reduced density matrix (populations and coherences)
%for the full GQME 
%were inaccurate 
%were found to be in reasonable qualitative agreement with the exact results, although clearly the noise and errors inherent to the NISQ era quantum machines negatively affected the accuracy~\cite{preskill2018quantum}; 2. the results for the GQME with reduced-dimensionality were found to be in quantitative agreements with the exact results, with significant improvements over the full GQME results. 
%These results clearly demonstrate that: 1. the quantum algorithm itself is accurate and the noises and errors observed in Fig.~\ref{fig:GQME_TFDTT_model_real} are entirely due to the hardware; 2. Dimensionality-reduction can greatly reduce the circuit depth and thus drastically improve the accuracy of simulations performed on real quantum computers. 

%There are several future directions of research that have the potential to improve the performance of the quantum algorithm on the NISQ quantum machines. One direction is to 

We acknowledge the fact that demonstrating quantum advantage is currently challenging for the proposed quantum algorithm used to simulate open quantum dynamics. However, quantum dynamics simulations often become computationally intractable on a classical computer even when the propagator is known (or numerically determined). This is simply because the time evolving state becomes highly entangled and therefore requires an exponentially large $N$ memory space, with $N$ the number of possible states in Hilbert space (and an exponentially large computational effort). In contrast, a quantum computer can efficiently represent the time-evolving state with only $\log_2(N)$ qubits. 
In addition, further improvement to our quantum simulations can be achieved by reducing the circuit depth via optimizing the quantum circuit design. 
%In general, the shallower the circuit depth (in terms of the number of elementary gates), the less the overall error in simulations on real quantum devices.  
%As discussed above, the simulations of the reduced-dimensionality GQMEs (see Fig.~\ref{fig:reduced}) clearly illustrate the notable improvement in the accuracy of the real machine simulation as a result of the shallower quantum circuit. 
%To this end, there are also many other proposals for possible ways to 
This can be achieved by optimizing the decomposition of unitary operations into elementary gate sequences~\cite{vartiainen2004efficient,gyongyosi2020quantum,lacroix2020improving,iten2022exact}. One particularly interesting idea is to reduce the circuit depth by adding qubits.\cite{abdessaied2013reducing} 
%which may be applicable to the current model, as 
To this end, it should be noted that we have only used 3 qubits out of the 5 currently available on the IBM quantum computers. Another way for improving accuracy is by active error correction using dynamical decoupling (DD) protocols,
%converts quantum gates into 
that employ 
%decoupling 
pulses 
%that can 
to suppress the system's coupling with the environment.\cite{viola1998dynamical,uhrig2007keeping,khodjasteh2005fault,khodjasteh2007performance,west2010near} 
Recent implementations of DD on IBM machines was found to improve  the fidelity of the overall performance~\cite{pokharel2018demonstration,das2021adapt,jurcevic2021demonstration}. 
%The error mitigation tools internally included by the IBM machines may also improve the overall simulation results and will be considered for the future implementations.     
Yet another direction is to implement the circuit on high-dimensional qudit machines. Quantum computers based on three-dimensional circuit quantum electrodynamics (3D cQED) microwave cavities are particularly promising in this respect, as they feature unique quantum error correction schemes \cite{ofek2016extending,hu2019quantum,campagne2020quantum} and longer coherence times \cite{paik2011observation,reagor2016quantum} than standard superconducting quantum computers. Bosonic quantum computing algorithms have also been recently shown to significantly reduce the number of quantum gates required for the calculation of the Franck-Condon factors  \cite{wang2020efficient} and dynamics of rhodopsin near conical intersections \cite{wang2022observation}. Lossless 3D cQED systems have not yet been employed to simulate open quantum system dynamics. An adaptation of the algorithm presented here to bosonic quantum computing could therefore provide another way to efficiently simulate open quantum system dynamics and demonstrate how qudit-based quantum architectures can reduce the computational cost and enhance the accuracy of quantum simulations.
%%%%%%%%%%%%%%%%%%%%%%%%%%%%%%%%%%%%%%%%
\newpage
\appendix

\section{Quantum circuit examples}
In this section, we include further details concerning the quantum algorithm, including the dilation process, circuit transpiling, QASM simulations, and simulations running on the IBM quantum computers \textbf{ibmq quito} and \textbf{ibmq lima}.
The normalized time evolution operator of the electronic reduced density operator ${\cal G}^{\prime}(t)={\cal G}(t)/n_c$ (where ${\cal G}(t)$ is generated from the GQME formalism) is dilated into a unitary operator ${\cal U}_{{\cal G}^{\prime}}$(t).
We start with ${\cal G}_3$, which corresponds to the ${\cal G}(t)$ of the $1500^{\mathrm{th}}$ time step from model 3, and ${\cal G}_4$, which corresponds to the ${\cal G}(t)$ of the $1500^{\mathrm{th}}$ time step from model 4.
The matrix of ${\cal G}_3$ and ${\cal G}_4$ are, respectively:
\begin{equation}
\resizebox{0.91\hsize}{!}{%
    ${\cal G}_3=\begin{pmatrix}
     0.38-\num{3.76e-10}j  &0.04+\num{2.90e-02}j & 0.04-\num{2.90e-02}j & 0.06-\num{1.88e-10}j\\
 -0.13+\num{7.04e-02}j  &0.28-\num{2.63e-02}j & 0.02+\num{2.37e-02}j &-0.15-\num{3.06e-02}j\\
 -0.13-\num{7.04e-02}j  &0.02-\num{2.37e-02}j  &0.28+\num{2.63e-02}j &-0.15+\num{3.06e-02}j\\
  0.62+\num{3.77e-10}j &-0.04-\num{2.90e-02}j &-0.04+\num{2.90e-02}j & 0.94+\num{1.87e-10}j
    \end{pmatrix},$%
    }
\end{equation}
and

\begin{equation}
    \resizebox{0.91\hsize}{!}{%
   $ {\cal G}_4=\begin{pmatrix}
    0.54+\num{4.7e-11}j &\num{-1.7e-06}+\num{5.7e-02}j &\num{-1.6e-06}-\num{5.6e-02}j&   0.46+\num{7.1e-11}j\\
 -0.46+\num{5.7e-02}j & \num{3.6e-02}+\num{6.1e-05}j& \num{-1.6e-02}-\num{5.7e-05}j&  -0.46-\num{5.7e-02}j\\
 -0.46-\num{5.7e-02}j& \num{-1.6e-02}+\num{5.7e-05}j & \num{3.7e-02}-\num{6.1e-05}j&  -0.46+\num{5.7e-02}j\\
  0.54-\num{4.7e-11}j & \num{1.6e-06}-\num{5.6e-02}j & \num{1.6e-06}+\num{5.6e-02}j&   0.54-\num{7.1e-11}j
    \end{pmatrix}.$%
    }
\end{equation}
The normalization factors used for model 3 and model 4 are $n_{c3}=1.376$ and  $n_{c4}=1.376$. 

\begin{figure}[h]
\centering
    \subfloat[${\cal U}_{{\cal G}_3}$ real part]{
	\begin{minipage}[t]{0.49\linewidth}
	   \centering
	   \includegraphics[width=1\textwidth]{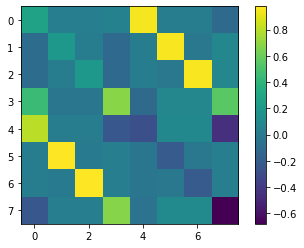}
	\end{minipage}}
 	\subfloat[${\cal U}_{{\cal G}_3}$ imaginary part]{
	\begin{minipage}[t]{0.49\linewidth}
	   \centering
	   \includegraphics[width=1\textwidth]{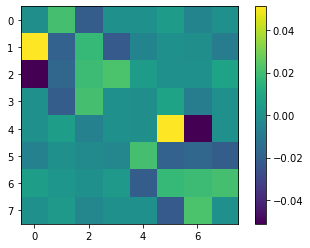}
	\end{minipage}}

  \subfloat[${\cal U}_{{\cal G}_4}$ real part]{
	\begin{minipage}[b]{0.49\linewidth}
	   \centering
	   \includegraphics[width=1\textwidth]{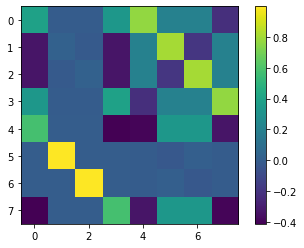}
	\end{minipage}}
    \subfloat[${\cal U}_{{\cal G}_4}$ imaginary part]{
	\begin{minipage}[b]{0.49\linewidth}
	   \centering
	   \includegraphics[width=1\textwidth]{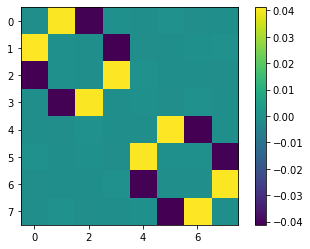}
	\end{minipage}}
    
\caption{Heat map illustrations of the dilated $8\times 8$ unitary matrix ${\cal U}_{{\cal G}_3}$ and ${\cal U}_{{\cal G}_4}$ for the $1500^{\mathrm{th}}$ ${\cal G}(t)$ matrix, ${\cal G}_3$ at $t=2.25\Gamma^{-1}$ for model 3 with (a) the real part of the matrix and (b) the imaginary part of the matrix and ${\cal G}_4$ at $t=6.75\Gamma^{-1}$ for model 4 with (c) the real part of the matrix and (d) the imaginary part of the matrix.}
	\label{fig:U_G}
\end{figure}

Following the $1$-dilation process, the $4\times 4$ ${\cal G}^{\prime}(t)$ [derived from corresponding ${\cal G}(t)$ divided by the $n_c$ factor] is converted into a unitary $8\times 8$ ${\cal U}_{{\cal G}^{\prime}}(t)$. We show  ${\cal U}_{{\cal G}^{\prime}_3}$ and ${\cal U}_{{\cal G}^{\prime}_4}$ in the form of heat maps in Fig.~\ref{fig:U_G}.

\begin{figure}[h]
\centering
    \subfloat[QASM simulation of model 3]{
	\begin{minipage}[t]{0.49\linewidth}
	   \centering
	   \includegraphics[width=1\textwidth]{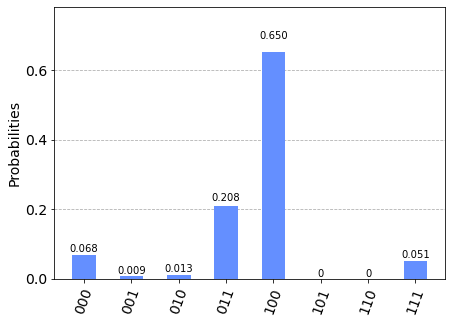}
	\end{minipage}}
 	\subfloat[\textbf{ibm quito} simulation of model 3]{
	\begin{minipage}[t]{0.49\linewidth}
	   \centering
	   \includegraphics[width=1\textwidth]{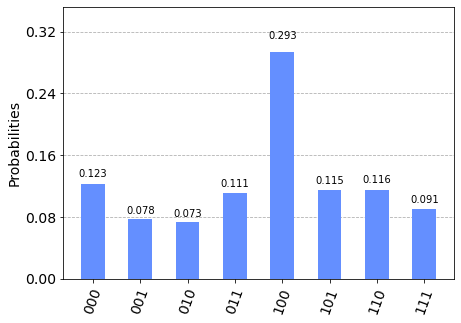}
	\end{minipage}}

  \subfloat[QASM simulation of model 4]{
	\begin{minipage}[b]{0.49\linewidth}
	   \centering
	   \includegraphics[width=1\textwidth]{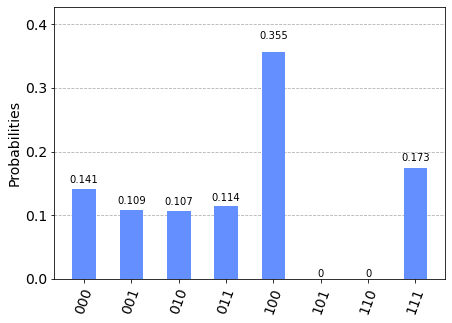}
	\end{minipage}}
    \subfloat[ \textbf{ibmq lima} simulation of model 3]{
	\begin{minipage}[b]{0.49\linewidth}
	   \centering
	   \includegraphics[width=1\textwidth]{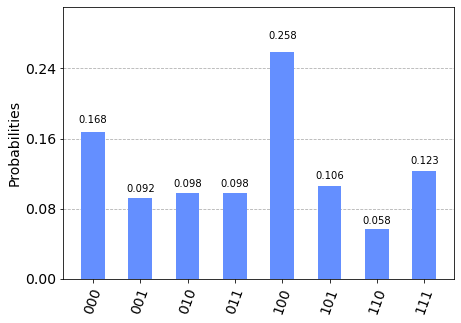}
	\end{minipage}}
    
\caption{Probability distribution of the quantum state after the projection measurement applied to the circuit for ${\cal U}_{{\cal G}_3}$ on (a) the QASM and (b) \textbf{ibm quito} quantum computer and ${\cal U}_{{\cal G}_4}$ on  (c) the QASM and (d) \textbf{ibm lima} quantum computer. The $|000\rangle$ state corresponds to the population squared of the donor state $\sigma_{DD}(t)$ and the $|100\rangle$ state corresponds to the population squared of the acceptor state $\sigma_{AA}(t)$. The last four states are ancilla states.}
	\label{fig:G_Prob_TFDTT_model1}
\end{figure}

\begin{figure}[h]
\includegraphics[width=10cm]{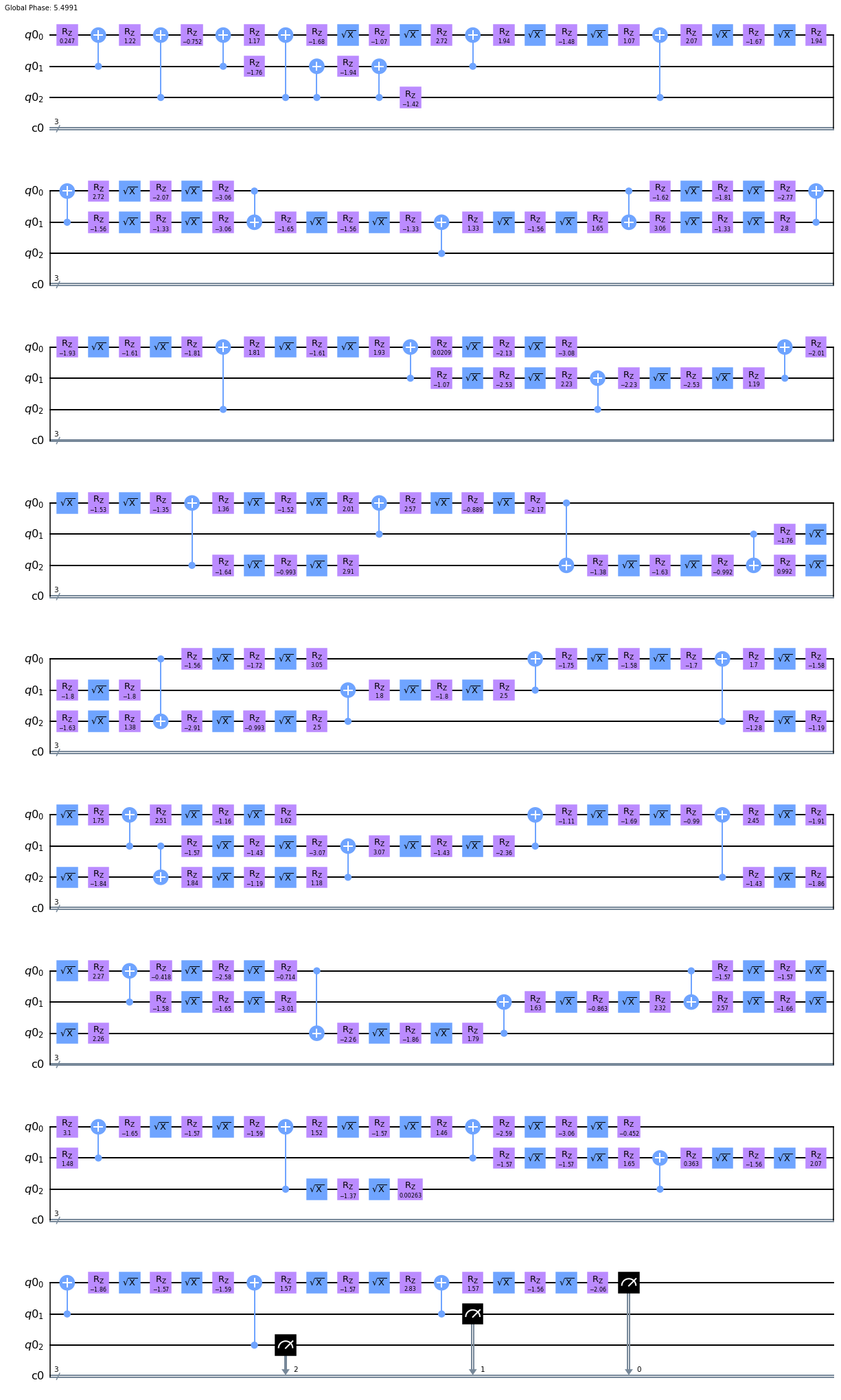}
\caption{Transpiled quantum circuit of the dilated ${\cal U}_{{\cal G}_3}$ matrix at $1500$ time steps for model 3. Each horizontal black line denotes a qubit. The $\sqrt{X}$ gate (blue square) is the square root of $X$ gate; the $R_z$ gate (magenta square) is the rotation $Z$ gate. The two-qubit gates are the controlled-NOT gate, where the dot denotes the controlled qubit and $\bigoplus$ denotes the target qubit. The black gates at the end of the circuit denote the projection measurements. The number of required $R_z$,  $\sqrt{X}$, and CNOT gates are $153,98$, and $41$, respectively.}
\label{fig:G_circuit_TFDTT_model4_1500}
\end{figure}

\begin{figure}[h]
\includegraphics[width=10cm]{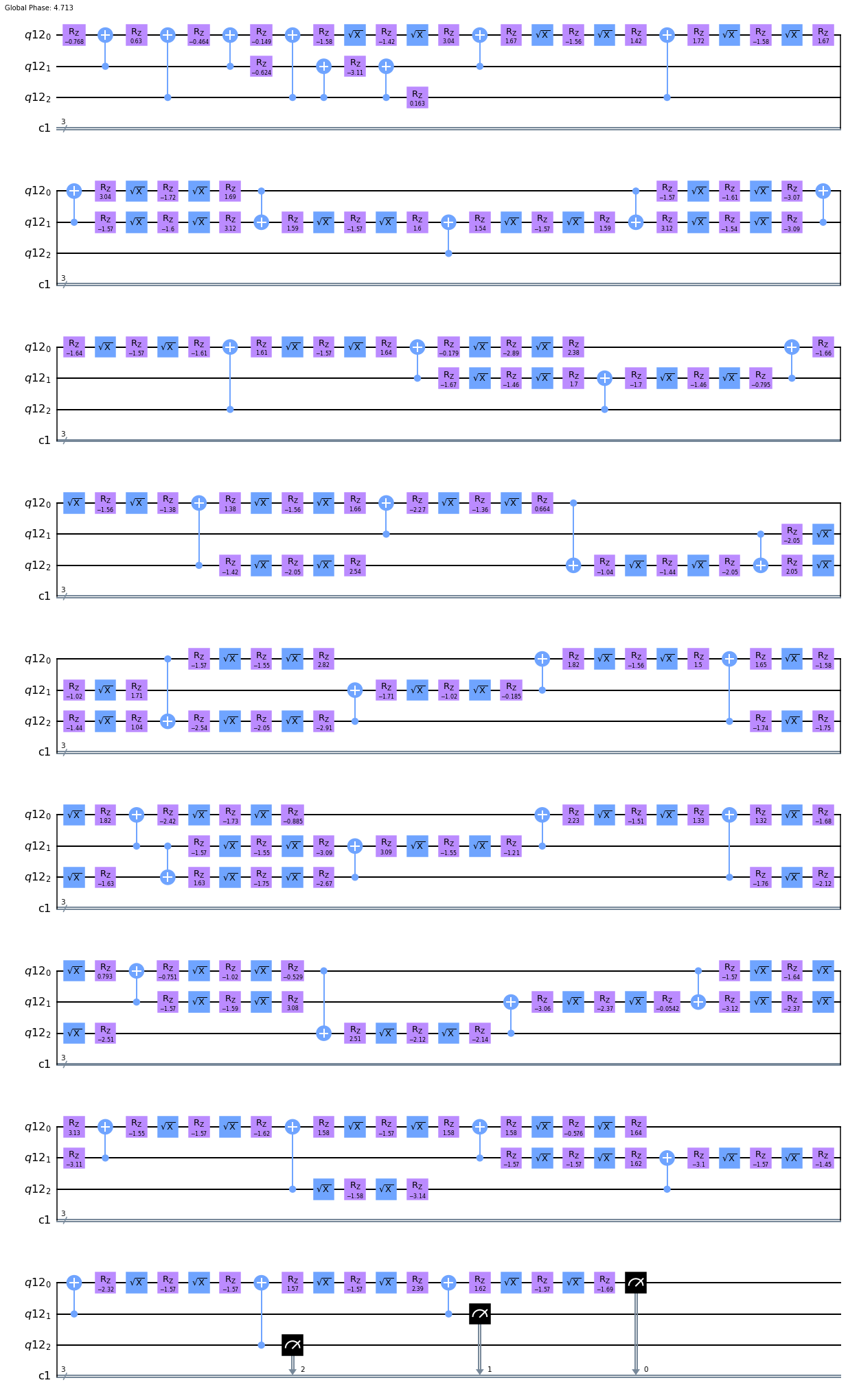}
\caption{Transpiled quantum circuit of the dilated ${\cal U}_{{\cal G}_4}$ matrix at $1500$ time steps for model 4. Each horizontal black line denotes a qubit. The $\sqrt{X}$ gate (blue square) is the square root of $X$ gate; the $R_z$ gate (magenta square) is the rotation $Z$ gate. The two-qubit gates are the controlled-NOT gate, where the dot denotes the controlled qubit and $\bigoplus$ denotes the target qubit. The black gates at the end of the circuit denote the projection measurements. The number of required $R_z$,  $\sqrt{X}$, and CNOT gates are $153,98$, and $41$, respectively.}
\label{fig:G_circuit_TFDTT_model6_1500}
\end{figure}

The unitary operation ${\cal U}_{{\cal G}^{\prime}}(t)$ is transpiled into a $3$-qubit quantum circuit composed of three elementary quantum gates: $R_Z$, $\sqrt{X}$, and $CX$, which have the matrix form:
\begin{align}
    \begin{split}R_Z(\lambda) &= \exp\left(-i\frac{\lambda}{2}Z\right) =
    \begin{pmatrix}
        e^{-i\frac{\lambda}{2}} & 0 \\
        0 & e^{i\frac{\lambda}{2}}
    \end{pmatrix}\end{split} \\
    \begin{split}\sqrt{X} &= \frac{1}{2} \begin{pmatrix}
        1 + i & 1 - i \\
        1 - i & 1 + i
    \end{pmatrix}\end{split} \\
    \begin{split}CX\ q_0, q_1 &=
    I \otimes |0\rangle\langle0| + X \otimes |1\rangle\langle1| =
    \begin{pmatrix}
        1 & 0 & 0 & 0 \\
        0 & 0 & 0 & 1 \\
        0 & 0 & 1 & 0 \\
        0 & 1 & 0 & 0
    \end{pmatrix}\end{split}
\end{align}
The full quantum circuits for ${\cal U}_{{\cal G}_3}$ and ${\cal U}_{{\cal G}_4}$ are shown in Fig.~\ref{fig:G_circuit_TFDTT_model4_1500} and \ref{fig:G_circuit_TFDTT_model6_1500}. The probability distribution of the projection measurement results of the two circuits are shown in Fig.~\ref{fig:G_Prob_TFDTT_model1}. Both the QASM simulator results and the real machine simulated results are recorded.

%%%%%%%%%%%%%%%%%%%%%%%%%%%%%%%%%%%%%%%%%%%%%%%%%%%%%%%%%%%%%%%%%%%%%%%%

\section{Amplitude-damping model}

%\textcolor{red}{I am not clear on the purpose of Secs.~A2 and A3, since they look like they are based on Lindblad QME and Kraus form? -- E. Geva}

%%%%%%%%%%%%%%%%%%%%%%%%%%%%%%%%%%%%%%%%%%%%%%%%%%%%%%%%%%%%%%%%%%%%%%%%
In this section, we will show that the method of flattening the density matrix for dilation, which was outlined in the supplementary information in our previous publication~\cite{hu2020quantum}, can be implemented on the quantum device for the simple amplitude damping model. The same method is verified by the implementation described in this and the following section. This verification allows us to confidently incorporate the general algorithm with GQME.

The general algorithm for open quantum system dynamics is applicable to the time-evolution of density matrices governed by Kraus operators~\cite{hu2020quantum}.
The time-evolution representation for such open systems is given by
$\hat{\rho}(t)=\sum\limits_{k}{{{\cal{M}}_{k}(t)}\hat{\rho} {{{\cal{M}}_{k}^{\dagger}}(t)}}$. For simplicity, the notation of time dependency is omitted hereafter for superoperators ${\cal{M}}$ and superoperators derived from it. 
In the first step, the density matrix is flattened to vector form: $\hat{\rho} \to {{\mathbf{v}}_{\rho }}={{\left( {{\rho }_{11}},...,{{\rho }_{1n}},{{\rho }_{21}},...,{{\rho }_{2n}},... \, ...,{{\rho }_{n1}},...,{{\rho }_{nn}} \right)}^{T}}$. We calculate the Frobenius norm of $\mathbf{v}_{\rho }$ as $\left\| {{\mathbf{v}}_{\rho }} \right\|_F=\sqrt{\sum\limits_{ij}{{{\left| {{\rho }_{ij}} \right|}^{2}}}}$ and divide $\mathbf{v}_{\rho }$ by $\left\| {{\mathbf{v}}_{\rho }} \right\|_F$ to normalize $\mathbf{v}_{\rho }$.
Next, for every $k$, the  ${{\cal{M}}_{k}}$ is transformed into ${{\cal {\tilde{M}}}_{k}}={{\cal{M}}_{k}}\otimes {I}$; similarly, the ${\cal{M}}_{k}^{\dagger}$ is transformed into ${{\cal {\tilde{N}}}_{k}}={I}\otimes {{\cal{\bar{M}}}_{k}}$. The $\otimes $ stands for the Kronecker product and the bar over ${{\cal{M}}_{k}}$ indicates complex conjugation. 
The new equivalent form for the Kraus  representation is:
\begin{equation}
    {{\cal{M}}_{k}}\hat \rho {\cal{M}}_{k}^{\dagger }\overset{\text{equivalent}}{\longleftrightarrow}{{\tilde{\cal N}}_{k}}{{\tilde{\cal M}}_{k}}{{\mathbf{v}}_{\rho }}.
\end{equation}
The input state is initialized to the normalized ${{\mathbf{v}}_{\rho }}$ in the execution. 
In Ref.~\citenum{hu2020quantum}, it is shown that the Kraus operator ${{\cal{M}}_{k}}$ is a contraction. Therefore ${{\tilde{\cal M}}_{k}}={{\cal{M}}_{k}}\otimes {I}$ and ${{\tilde{\cal N}}_{k}}={I}\otimes {{\cal{\bar{M}}}_{k}}$ are also contractions as per the norm property of the Kronecker product.
To build the quantum circuit of ${{\tilde{\cal N}}_{k}}{{\tilde{\cal M}}_{k}}{{\mathbf{v}}_{\rho }}$ with unitary gates, we need two separate 2-dilations:
\begin{equation}
    {{ \tilde{\cal N}}_{k}}{{\tilde{\cal M}}_{k}}{{\mathbf{v}}_{\rho }}\xrightarrow{\text{unitary dilation}}{{\mathbf{\cal{U}}}_{{{\cal{N}}_{k}}}{\mathbf{\cal{U}}}_{{{\cal{M}}_{k}}}}{{\left( \mathbf{v}_{\rho }^{T},0,...,0 \right)}^{T}}.
\end{equation}
For ${{\cal{M}}_{k}}$ of dimension $n \times n$, ${{\tilde{\cal M}}_{k}}$ and ${{\tilde{\cal N}}_{k}}$ are ${{n}^{2}} \times {{n}^{2}}$; and consequently, the 2-dilations ${{\mathbf{\cal{U}}}_{{{\cal M}_{k}}}}$ and ${{\mathbf{\cal U}}_{{{\cal N}_{k}}}}$ are $3{{n}^{2}} \times 3{{n}^{2}}$.
The ${{\mathbf{\cal U}}_{{{\cal M}_{k}}}}$ and ${{\mathbf{\cal U}}_{{{\cal N}_{k}}}}$ are fragmented into sequences of two-level unitary gates and tallied to compute the gate complexity. To realize ${{\tilde{\cal N}}_{k}}{{\tilde{\cal M}}_{k}}{{\mathbf{v}}_{\rho }}$, the total gate complexity is $3{{n}^{3}}+{{n}^{2}}$ for each $k$. The classical complexity to realize ${{{\cal M}}_{k}}\hat{\rho} {\cal M}_{k}^{\dagger }$ based on a naive algorithm is higher, though of same order as quantum algorithm, namely $4{{n}^{3}}-2{{n}^{2}}$. 

All the evolved density matrices in the circuit calculated at each timestep are obtained as the output vector ${{\mathbf{v}}_{k}}\left( t \right)={{\tilde{\cal N}}_{k}}{{\tilde{\cal M}}_{k}}{{\mathbf{v}}_{\rho }}$. The desired information to be collected from the density matrix is extracted by applying projection measurements on  ${{\mathbf{v}}_{k}}\left( t \right)$ using an optical setup~\cite{sparrow2018simulating}.
%\textbf{Ref.~1.} 
The detailed procedure for obtaining information located at both diagonal and off-diagonal elements of ${{\hat{\rho} }_{k}}\left( t \right)$ from final ${{\mathbf{v}}_{k}}\left( t \right)$ is described in the supplementary information of Ref.~\citenum{hu2020quantum}. 

\section{Simulation of the amplitude damping model with Kraus operators}

We tested the theory mentioned in the previous section for spontaneous emission of a 2-level atom modeled by amplitude-channel damping. The corresponding Lindblad master equation is: 
$${{\mathbf{\dot{\hat \rho}}}}\left( t \right)={\gamma} \left[{\sigma }^{+}\hat{\rho} (t){\sigma }^{-}-\frac{1}{2}\{{\sigma }^{-}{\sigma }^{+},\hat{\rho} (t)\}\right],$$ 
where the spontaneous emission rate is $\gamma =1.52\times {{10}^{9}}\;{{ \text{s}}^{-1}}$, and the ${\sigma }^{+}=|0\rangle \langle 1|$ and ${\sigma }^{-}={({\sigma }^{+})}^{\dagger }$ are Pauli raising and lowering operators, respectively. The density matrix $\rho (t)$ in the Kraus representation is as follows:
\begin{align}
    \begin{array}{ccc}\hat{\rho} (t) & = & {{\cal{M}}}_{0}(t)\hat{\rho} {{\cal{M}}}_{0}{(t)}^{\dagger }+{{\cal{M}}}_{1}{(t)}\hat \rho {{\cal{M}}}_{1}{(t)}^{\dagger },\\ {{\cal{M}}}_{0}{(t)} & = & \frac{1+\sqrt{{e}^{-\gamma t}}}{2}{\bf{I}}+\frac{1-\sqrt{{e}^{-\gamma t}}}{2}{\sigma }_{z}=(\begin{array}{cc}1 & 0\\ 0 & \sqrt{{e}^{-\gamma t}}\end{array}),\\ {{\cal{M}}}_{1}{(t)} & = & \sqrt{1-{e}^{-\gamma t}}{\sigma }^{+}=(\begin{array}{cc}0 & \sqrt{1-{e}^{-\gamma t}}\\ 0 & 0\end{array}).\end{array}
\end{align}

For ${{{\cal M}}_{k}}$ of dimension $2 \times 2$, ${{\tilde {\cal M}}_{k}}$, ${{\tilde{\cal N}}_{k}}$, and ${\cal D}_{A}$ are $4 \times 4$ matrices, as given below in Eq.~\eqref{eq:MNDmats}. In this way, the 2-dilations ${{\mathbf{\cal U}}_{{{\cal M}_{k}}}}$ and ${{\mathbf{\cal U}}_{{{\cal N}_{k}}}}$ are $12 \times 12$ following the $k$-dilation~\cite{hu2020quantum,levy2014dilation}. Note that, though these superoperators are time dependent, only for simplicity we omitted the notation of time dependency. However, realization of the dilated matrices using quantum gate is of dimension of form ${2}^{n} \times {2}^{n}$. We append the dilated matrix with an ancillary $12 \times 4$ zero matrix on the right and $ 4 \times 12$ at the bottom, and an $4 \times 4$ identity matrix along the diagonal. The resulting dilated superoperator matrix is $16 \times 16$, requiring 4 qubits for quantum implementation. Quantum implementation is accomplished with Qiskit as mentioned in the main text. 

\begin{align}
  & {{\tilde{\cal M}}_{0}}=\left( \begin{matrix}
  1 & 0 & 0 & 0  \\
  0 & 1 & 0 & 0  \\
  0 & 0 & \sqrt{{{e}^{-\gamma t}}} & 0  \\
  0 & 0 & 0 & \sqrt{{{e}^{-\gamma t}}}  \\
\end{matrix} \right),\quad {{\tilde{\cal N}}_{0}}=\left( \begin{matrix}
  1 & 0 & 0 & 0  \\
  0 & \sqrt{{{e}^{-\gamma t}}} & 0 & 0  \\
  0 & 0 & 1 & 0  \\
  0 & 0 & 0 & \sqrt{{{e}^{-\gamma t}}}  \\
\end{matrix} \right),\quad \nonumber\\ 
 & {{\mathbf{\cal D}}_{{{\tilde{\cal M}}_{0}}}}=\left( \begin{matrix}
  0 & 0 & 0 & 0  \\
  0 & 0 & 0 & 0  \\
  0 & 0 & \sqrt{1-{{e}^{-\gamma t}}} & 0  \\
  0 & 0 & 0 & \sqrt{1-{{e}^{-\gamma t}}}  \\
\end{matrix} \right),\quad {{\mathbf{\cal D}}_{{{\tilde{\cal N}}_{0}}}}=\left( \begin{matrix}
  0 & 0 & 0 & 0  \\
  0 & \sqrt{1-{{e}^{-\gamma t}}} & 0 & 0  \\
  0 & 0 & 0 & 0  \\
  0 & 0 & 0 & \sqrt{1-{{e}^{-\gamma t}}}  \\
\end{matrix} \right).       \nonumber \\
  & {{\tilde{\cal M}}_{1}}=\left( \begin{matrix}
  0 & 0 & \sqrt{1-{{e}^{-\gamma t}}} & 0  \\
  0 & 0 & 0 & \sqrt{1-{{e}^{-\gamma t}}}  \\
  0 & 0 & 0 & 0  \\
  0 & 0 & 0 & 0  \\
\end{matrix} \right),\text{                 }{{\tilde{\cal N}}_{1}}=\left( \begin{matrix}
  0 & \sqrt{1-{{e}^{-\gamma t}}} & 0 & 0  \\
  0 & 0 & 0 & 0  \\
  0 & 0 & 0 & \sqrt{1-{{e}^{-\gamma t}}}  \\
  0 & 0 & 0 & 0  \\
\end{matrix} \right),\text{  } \nonumber\\ 
 & {{\mathbf{\cal D}}_{{{\tilde{\cal M}}_{1}}}}=\left( \begin{matrix}
  1 & 0 & 0 & 0  \\
  0 & 1 & 0 & 0  \\
  0 & 0 & \sqrt{{{e}^{-\gamma t}}} & 0  \\
  0 & 0 & 0 & \sqrt{{{e}^{-\gamma t}}}  \\
\end{matrix} \right),\text{                        }{{\mathbf{\cal D}}_{{{\tilde{\cal N}}_{1}}}}=\left( \begin{matrix}
  1 & 0 & 0 & 0  \\
  0 & \sqrt{{{e}^{-\gamma t}}} & 0 & 0  \\
  0 & 0 & 1 & 0  \\
  0 & 0 & 0 & \sqrt{{{e}^{-\gamma t}}}  \\
\end{matrix} \right) . \nonumber \\  \label{eq:MNDmats}
\end{align} 	%(14)
 
\begin{figure}[h]
\centering
\includegraphics[width=8cm]{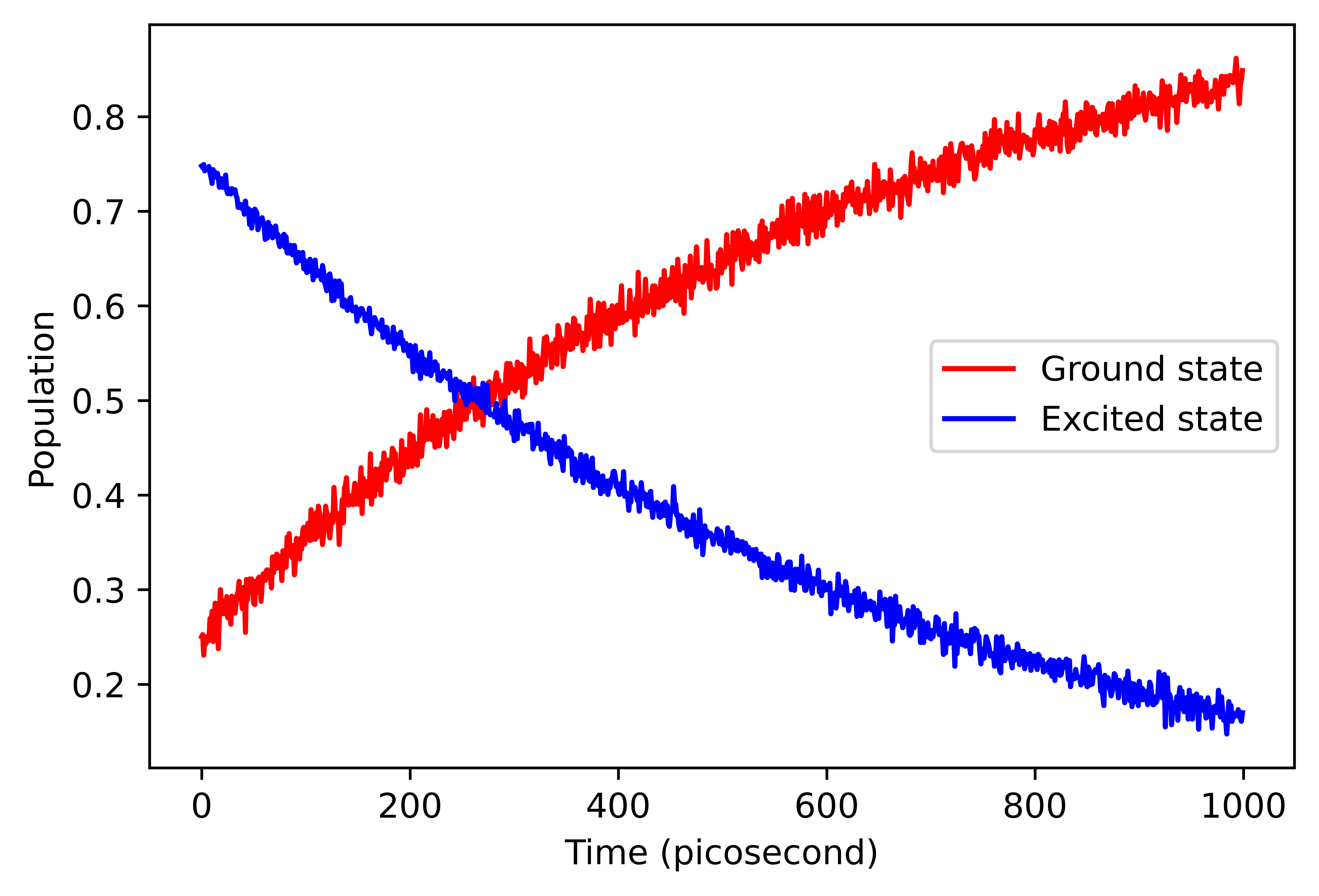}
\caption{ Population of ground state and excited state for the amplitude-damping model obtained by the quantum implementation on the IBM Qiskit simulator.}
\label{fig:amplitude-damping QASM}
	%\label{fig:QASM_simulation}https://www.overleaf.com/project/628f957bfcaddf9ce211a295
\end{figure}

For an initial density $\hat\rho(0) =\frac{1}{4}\left( \begin{matrix}
   1 & 1  \\
   1 & 3  \\
\end{matrix} \right)$, we calculate the populations in the basis $\{|0\rangle ,|1\rangle \}$ from $t=0$ to $t=1000\text{ ps}$ with a time step of 10 ps. With ${{\left\| \hat{\rho}  \right\|}_{HS}}=\frac{\sqrt{3}}{2}$, the input state is:
\begin{equation}
    {{\mathbf{v}}_{0}}=\frac{1}{{{\left\| \hat \rho  \right\|}_{HS}}}{{\left( \mathbf{v}_{\rho }^{T},\overbrace{0,...,0}^{m} \right)}^{T}}=\frac{1}{2\sqrt{3}}{{\left( 1,1,1,3,\overbrace{0,...,0}^{m} \right)}^{T}},
\end{equation}
where $m=12$ for the vector $\mathbf{v}_{\rho }^{T}$ to be of length 16. After extracting the output ${{\mathbf{v}}_{k}}\left( t \right)$, the ground state and excited state populations are obtained as the first and fourth entry of the vector, respectively. The Fig.~\ref{fig:amplitude-damping QASM} result manifests the consistency with the result in Ref.~\citenum{hu2020quantum}. 

\begin{acknowledgement}

We acknowledge the financial support of  the National Science Foundation under award number 2124511, CCI Phase I: NSF Center for Quantum Dynamics on Modular Quantum Devices (CQD-MQD). We acknowledge the use of IBM Quantum services for this work.
The views expressed are those of the authors and do not
reflect the official policy or position of IBM or the IBM
Quantum team.

\end{acknowledgement}


\bibliography{References}

\end{document}